\documentclass[reprint,amsmath,amsfonts,amssymb,prb,aps,longbibliography,nofootinbib]{revtex4-1}
\usepackage{graphicx}
\usepackage{siunitx}
\usepackage{braket}
\usepackage{xcolor}
\usepackage[normalem]{ulem}
\newcommand{\subheading}[1]{\vspace{10pt}\noindent\textbf{#1}\\[3pt]}

\begin{document}

	\title{Deterministic multi-mode gates on a scalable photonic quantum computing platform}
	\author{Mikkel V. Larsen}
	\email{mikkel.vilsboell@gmail.com}
	\author{Xueshi Guo}
	\author{Casper R. Breum}
	\author{Jonas S. Neergaard-Nielsen}
	\author{Ulrik L. Andersen}
	\email{ulrik.andersen@fysik.dtu.dk}
	\affiliation{Center for Macroscopic Quantum States (bigQ), Department of Physics, Technical University of Denmark, Fysikvej, 2800 Kgs. Lyngby, Denmark}
	\date{October 27, 2020}
	\maketitle

\textbf{Quantum computing can be realized with numerous different hardware platforms and computational protocols. A highly promising approach to foster scalability is to apply a photonic platform combined with a measurement-induced quantum information processing protocol where gate operations are realized through optical measurements on a multipartite entangled quantum state---a so-called cluster state~\cite{raussendorf01,menicucci06}. Heretofore, a few quantum gates on non-universal or non-scalable cluster states have been realized~\cite{ukai11a,ukai11b,su13,asavanant20,Reimer2019,Gao2011,Walther2005,Lanyon2013}, but a full set of gates for universal scalable quantum computing has not been realized. We propose and demonstrate the deterministic implementation of a multi-mode set of measurement-induced quantum gates in a large two-dimensional (2D) optical cluster state using phase-controlled continuous variable quadrature measurements \cite{menicucci06,gu09}. Each gate is simply programmed into the phases of the high-efficiency quadrature measurements which execute the transformations by teleportation through the cluster state. Using these programmable gates, we demonstrate a small quantum circuit consisting of 10 single-mode gates and 2 two-mode gates on a three-mode input state. On this platform, fault-tolerant universal quantum computing is possible if the cluster state entanglement is improved and a supply of Gottesman-Kitaev-Preskill qubits is available\cite{gottesman01,baragiola19,yamasaki19,hastrup20}. Moreover, it operates at the telecom wavelength and is therefore network connectable without quantum transducers.}

Recent remarkable advances in developing fully programmable quantum computing platforms have led to a plethora of groundbreaking results in quantum information science including the demonstration of fault-tolerant operations on an error-corrected logical ion-trap qubit \cite{egan20} and the demonstration of quantum sampling at a super-classical rate in a 53-qubit superconducting quantum computer \cite{Arute2019}. Albeit marked progress, the currently realized qubit-based platforms for quantum computing are still strongly limited in size while the proposed methods for up-scaling are stymied by significant technical challenges. 

An alternative is the continuous variable (CV) photonic platform which has recently gained interest due to its proven scalability potential for measurement-based quantum computation (MBQC) as exemplified by the generation of 2D cluster states with thousands of modes~\cite{larsen19,asavanant19} and the sequential operation of one hundred single-mode gates~\cite{asavanant20}. In CV quantum computing~\cite{lloyd99,menicucci06,lund08,gottesman01}, information is encoded and processed in bosonic harmonic oscillators---e.g. the optical field---that are described by states in infinite-dimensional Hilbert spaces~\cite{Weedbrook2012,andersen15}. Although the idea of using CVs for quantum computing dates back more than 20 years \cite{lloyd99}, it is only within the last few years that feasible models for fault-tolerant large-scale CV MBQC were conceived~\cite{menicucci14,baragiola19,yamasaki19,fukui18,noh20}. Our demonstration represents a critical step towards these CV computing models. It constitutes the first realization of a fully deterministic and programmable multi-mode computation platform for MBQC.  

\begin{figure*}
	\includegraphics[width=0.9\linewidth]{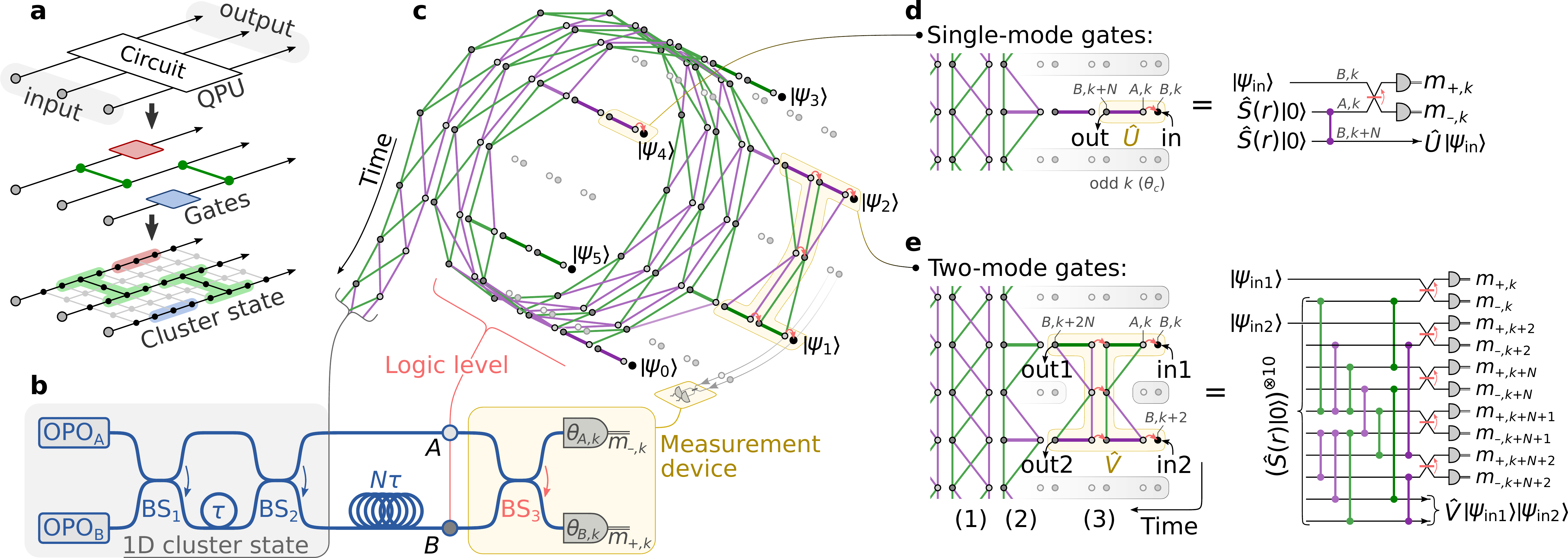}
	\caption{\label{fig:scheme}
	\textbf{Experimental setup and computation scheme.} \textbf{a}, Decomposition of a quantum circuit in a quantum processing unit (QPU) into gates implemented on a cluster state by projective measurements of the input and cluster state modes. \textbf{b}, Experimental setup generating a coiled-up 1D cluster state in the logic level where quantum information is encoded \cite{larsen20}. Computation takes place in the logic level using a two-mode measurement device consisting of a beam-splitter ($\text{BS}_3$) and two homodyne detectors	measuring in bases $\theta_{A,k}$ and $\theta_{B,k}$, with $k$ indicating the temporal mode number.	The experimental setup is described in detail in SI section 1. \textbf{c}, Coiled-up 1D cluster state in the logic level at which input states, $\ket{\psi_0},\dots,\ket{\psi_5}$, can be encoded on the circumference. By measuring control modes, the cluster state is projected into wires on which single- and two-mode gates can be implemented by gate teleportation. Bright and dark nodes indicate spatial modes $A$ and $B$ respectively, while the red arrows indicate $\text{BS}_3$-operation of the measurement device. \textbf{d} and \textbf{e}, Cut-outs of the cluster state showing implementations of single- and two-mode gate operations, $\hat{U}$ and $\hat{V}$, defined by the measurement device's basis settings. Here the coiled-up cluster state at (1) is seen to be projected into wires at (2) by measuring odd temporal control modes (grey shaded area) in the $\theta_c$ control basis prior to the implementation of gates at (3). The corresponding circuits are shown as well and are further described in SI section 2 together with the computation scheme.}
\end{figure*}

\subheading{Architecture and hardware}
In CV MBQC, quantum information processing is realized by teleporting the constituent gates through a computationally universal cluster state, comprising quadrature entangled modes in a 2D grid \cite{menicucci06}. For Gaussian gates, these teleportation protocols are effectuated by quadrature measurements of the cluster state modes, where the determined gate or sequence of gates is fully programmed into the phases of the quadrature measurements, i.e. the measurement bases. Such a cluster state architecture including solely Gaussian transformations allows for full computational universality in the sub-space of Gottesman-Kitaev-Preskill (GKP) qubits provided that a supply of these qubits are available \cite{gottesman01,baragiola19}---inline non-Gaussian gates are not needed\cite{baragiola19}. Moreover, by using GKP encoding, fault-tolerance is attainable via quantum error correction involving solely Gaussian transformations in the cluster\cite{gottesman01,walshe20}.

The reconfigurable and programmable nature of the cluster state quantum computer is illustrated in Fig.~\ref{fig:scheme}a by the different layers of operation from software to hardware. First, the quantum algorithm is specified and subsequently resolved into a certain sequence of single- and two-mode gates, such as the rotation, shear, squeezing and the controlled-Z gate. This sequence of gates is then converted into a sequence of phases that finally controls the consecutive quadrature measurements to effect the quantum algorithm on the cluster state. 

At the hardware level, the processor is based on the generation of a time-encoded one-dimensional (1D) cluster state \cite{menicucci11b,yokoyama13} as illustrated in Fig.~\ref{fig:scheme}b. The cluster state is created in two spatial modes, $A$ and $B$, using squeezed states of light (produced by optical parametric oscillators, $\text{OPO}_A$ and $\text{OPO}_B$, at the wavelength of $\SI{1550}{nm}$) that are interfered in an imbalanced interferometer and subsequently coiled up by the $N\tau$-delay into a cylindrical 2D cluster state with a structure as shown in Fig.~\ref{fig:scheme}c \cite{larsen19}. This coiled-up 1D cluster state represents the logic level, and it is at this point the computational logic takes place via projective quadrature measurements of the cluster state. Each node of the graph corresponds to spatial modes $A$ and $B$ in different temporal modes, $k$, with duration $\tau$ set by the $\tau$-delay in the 1D cluster state generator, and the number, $N$, of temporal modes on the cylinder circumference is determined by the $N\tau$-delay. We have chosen the delays of $\SI{50}{m}$ and $\SI{600}{m}$ leading to a temporal mode duration of $\tau\approx\SI{250}{ns}$ and $N=12$. The projective quadrature measurements of each node are performed with high-efficiency homodyne detectors with variable and fully controllable basis settings. It is precisely this full control of projective quadrature measurements that enables the execution of an arbitrary sequence of Gaussian gates and thus allows for universal quantum computation when using GKP states.

\subheading{Computation scheme}
Input quantum states for computation may be encoded on the circumference of the cluster state, and for the implementation of the desired quantum gates, they are teleported along the cylinder by projective measurement of each mode. The actually implemented gate depends on the measurement bases used for the teleportation. Note that the measurements are performed chronologically, swirling round the cylinder. This is exactly the right measurement order for universal computation on GKP-qubits---no additional optical storage is required since the measurement order follows the propagation of information along the cluster state cylinder. In the following, we summarize the computation scheme with details described in the supplementary information (SI) section 2.

At the logic level in Fig.~\ref{fig:scheme}b and 1c, a joint measurement is performed on spatial modes $A$ and $B$ of every temporal mode $k$ (examples of the mode indexing is shown in Fig.~\ref{fig:scheme}d,e). The two-mode measurement device consists of a beam-splitter ($\text{BS}_3$) followed by two homodyne detectors each measuring in a basis determined by the local oscillator phase $\theta$---i.e. measuring the quadrature $\hat{x}(\theta)=\hat{x}\cos\theta+\hat{p}\sin\theta$ where $\hat{x}$ and $\hat{p}$ are the electric field amplitude and phase (or position and momentum) quadratures, respectively. Temporal modes of odd $k$ are used as control modes: Measuring these in basis $\theta_c=(-1)^{(k-1)/2}\pi/4$ of $A$ and $B$, the cluster state is projected into $N/2$ wires along the length of the cylinder as illustrated in Fig.~\ref{fig:scheme}c \cite{menicucci11a}.  These wires, which at the logic level consist of segments of two-mode entangled states, can be used for single-mode computation by gate teleportation \cite{alexander14}. By performing a joint measurement on mode $B,k$ containing an input state and mode $A,k$ of the neighbouring wire segment, the input state is teleported through a gate, $\hat{U}$, to mode $B,k+N$ of the same wire. The operation $\hat{U}$ depends on the measurement basis setting, $(\theta_{A,k},\theta_{B,k})_U$ (see Fig.~\ref{fig:scheme}d). To implement two-mode gates, modes of neighboring wires must be coupled which can be done by changing the measurement basis, $\theta_c$, of some of the control modes \cite{larsen20,alexander16b}. Depending on the basis setting on the wire modes (indices $k, k+N$ for one wire and $k+2, k+N+2$ for the other) and the coupling control mode (index $k+N+1$), a two-mode gate operation $\hat{V}$ is implemented as illustrated in Fig.~\ref{fig:scheme}e.

In the Heisenberg picture, an implemented Gaussian $n$-mode gate operation transforms the quadratures as
\begin{equation}\label{eq:operation}
    \boldsymbol{\hat{q}'}=\textbf{G}\boldsymbol{\hat{q}}+\textbf{N}\boldsymbol{\hat{p}_i}+\textbf{D}\boldsymbol{m}\;,
\end{equation}
where $\boldsymbol{\hat{q}}=(\hat{x}_1,\cdots,\hat{x}_n,\hat{p}_1,\cdots,\hat{p}_n)^T$ and $\boldsymbol{\hat{q}'}=(\hat{x}'_1,\cdots,\hat{x}'_n,\hat{p}'_1,\cdots,\hat{p}'_n)^T$ are $2n$ vectors of quadratures of the gate input and output modes, respectively. For single- and two-mode gates, $n=1$ and $2$. In Eq.~\eqref{eq:operation}, the first term represents the Gaussian gate with $\textbf{G}$ being the corresponding symplectic matrix that depends on the measurement basis setting. The last term, $\textbf{D}\boldsymbol{m}$, represents a teleportation by-product of displacements with $\boldsymbol{m}$ being a vector of measurement outcomes, transformed by the basis setting dependent matrix $\textbf{D}$. Finally, the middle term, $\textbf{N}\boldsymbol{\hat{p}_i}$, represents noise occurring in the gate with $\boldsymbol{\hat{p}_i}$ being a vector of initial momentum squeezed quadratures of the cluster state modes, transformed by the gate noise matrix $\textbf{N}$. $\textbf{G}$, $\textbf{N}$ and $\textbf{D}$ are given in the SI section 2 for different basis settings.

An ideal gate transformation is performed when the gate noise and displacement terms are zero. Since the measurement outcomes, $\boldsymbol{m}$, are known, the displacement by-product can be compensated for by adding $-\textbf{D}\boldsymbol{m}$ to the measurement outcomes of output state quadratures, $\boldsymbol{\hat{q}'}$. The gate noise, however, is only negligible for cluster states generated from infinitely squeezed vacuum states, i.e.~$\text{Var}\{\boldsymbol{\hat{p}_i}\}=\boldsymbol{0}$. Such states are however nonphysical and in practice, non-zero additive Gaussian noise will inevitably occur and must eventually be accounted for by quantum error-correction.

Our proposed computational scheme of Gaussian gates offer full universality and fault-tolerance when provided with a supply of GKP qubits. Moreover, the scheme requires limited active feedforward: The displacement by-products, $\textbf{D}\boldsymbol{m}$, as well as the displacements needed for the noise correction, can be handled by post-processing of the measurement outcomes. The only active feedforward required is simple binary updating of the basis settings in later measurements to implement either a single-mode identity or a shear gate for each non-Clifford $\hat{T}$ gate\cite{gottesman01}.

\subheading{Quantum gates}
For single-mode gates, the implemented operation corresponding to \textbf{G} in Eq.~\eqref{eq:operation} is
\begin{equation}\label{eq:single-mode_gate}
    (-1)^w\hat{R}(\theta_+/2)\hat{S}(\tan\theta_-/2)\hat{R}(\theta_+/2)\;,
\end{equation}
where $w=(k\mod N)/2$ is the wire number, which in our realization runs from 0 to 5, $\hat{R}(\theta)=e^{i\theta(\hat{x}^2+\hat{p}^2)/2}$ and $\hat{S}(s)=e^{i\ln(s)(\hat{x}\hat{p}+\hat{p}\hat{x})}$ are rotation and squeezing operations, and $\theta_\pm$ depends on the basis settings as $\theta_\pm=\pm\theta_{A,k}+\theta_{B,k}$. We experimentally implemented the rotation gate, $\hat{R}(\theta)$, a modified shear gate, $\hat{F}^j\hat{P}(\sigma)=\hat{F}^je^{i\sigma\hat{x}^2/2}$, and the squeezing gate, $\hat{S}(e^r)$, by appropriately choosing the basis settings, as detailed in the SI section 2.1. Note, this modified version of the shear gate, with $\hat{F}^j=\hat{R}(j\pi/2)$ where $j=(-1)^w$, makes it possible to implement shear in a single computation step of Eq.~\eqref{eq:single-mode_gate}, while $\hat{F}^j$ can be compensated for in a second computation step if necessary. $\{\hat{R}(\theta),\hat{S}(e^r)\}$ constitutes a universal single-mode Gaussian gate-set \cite{ukai10}, while $\{\hat{R}(\pi/2),\hat{F}^j\hat{P}(1)\}$ constitutes a single-mode Clifford gate-set on GKP-encoded qubits \cite{gottesman01}. Note that in CV MBQC, similar to feedforward of displacement by-products and GKP noise correction, displacement gates for universal computation are implemented in the post-processing of the measurement outcomes \cite{menicucci06,gu09}.

\begin{figure*}
	\includegraphics[width=0.75\linewidth]{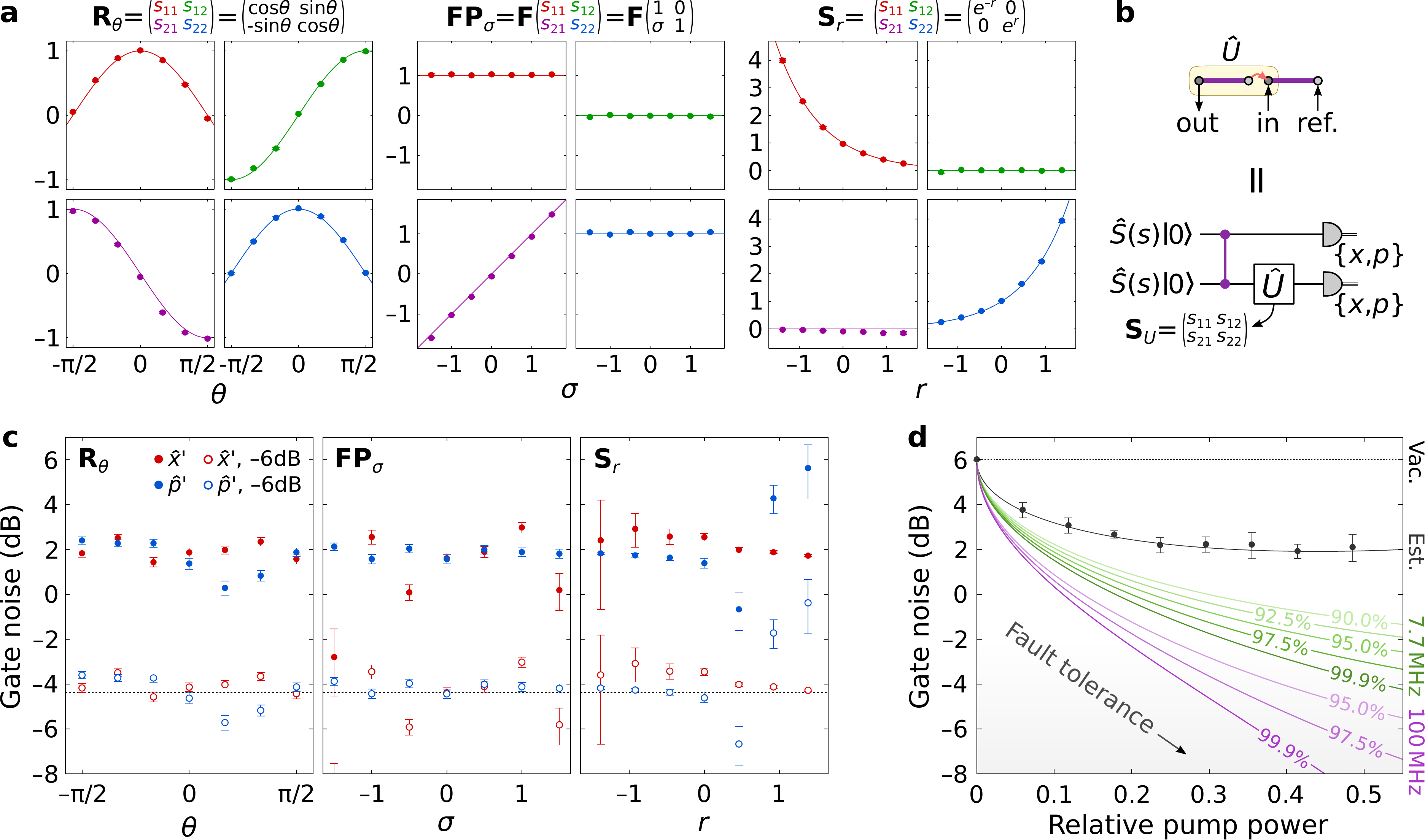}
	\caption{\label{fig:single_mode}
	\textbf{Single-mode gates.} \textbf{a}, Symplectic matrices corresponding to $\textbf{G}$ in Eq.~\eqref{eq:operation} for the implemented rotation, shear (modified by $\hat{F}$), and squeezing gates, measured by gate tomography as described in SI section 3 and summarized in \textbf{b}. \textbf{c}, Measured gate noise variance in each output quadrature for each implemented gate (solid points). Also, gate noise compensated by $\SI{-6}{dB}$ to account for the effect of the gate noise matrix $\textbf{N}$ in Eq.~\eqref{eq:operation} is shown (hollow points), in which case it should agree with the initial momentum squeezing variance, which is independently measured to be $\SI{-4.4}{dB}$ (dashed line). Each point in \textbf{a} and \textbf{c} is extracted from $\SI{10000}{}$ measurements, while the error bars are estimated as standard deviation when binning the $\SI{10000}{}$ measurements into smaller datasets. The same is the case for datapoints in Fig.~\ref{fig:CZ_gate} and \ref{fig:circuit}. \textbf{d}, Single-mode gate noise as a function of pump power, where zero pump power corresponds to no squeezing in which case the cluster state is replaced by vacuum. Here the measured gate noise (solid points) is for $\hat{R}(\theta)$, averaged over $\theta$, $\hat{x}$ and $\hat{p}$, and error bars are estimated as standard deviation hereof. Black-solid line represents the estimated gate noise using the setup parameters, while the dashed line corresponds to the case with no correlations. Green and purple lines show the expected gate noise for more optimal parameters with $\geq90\%$ optical efficiencies and for $\SI{7.7}{MHz}$ squeezing bandwidth as in the experimental setup (green), together with a broader squeezing bandwidth of $\SI{100}{MHz}$ (purple)---see SI section 4 for a derivation and further discussion.}
\end{figure*}

\begin{figure*}
	\includegraphics[width=0.56\linewidth]{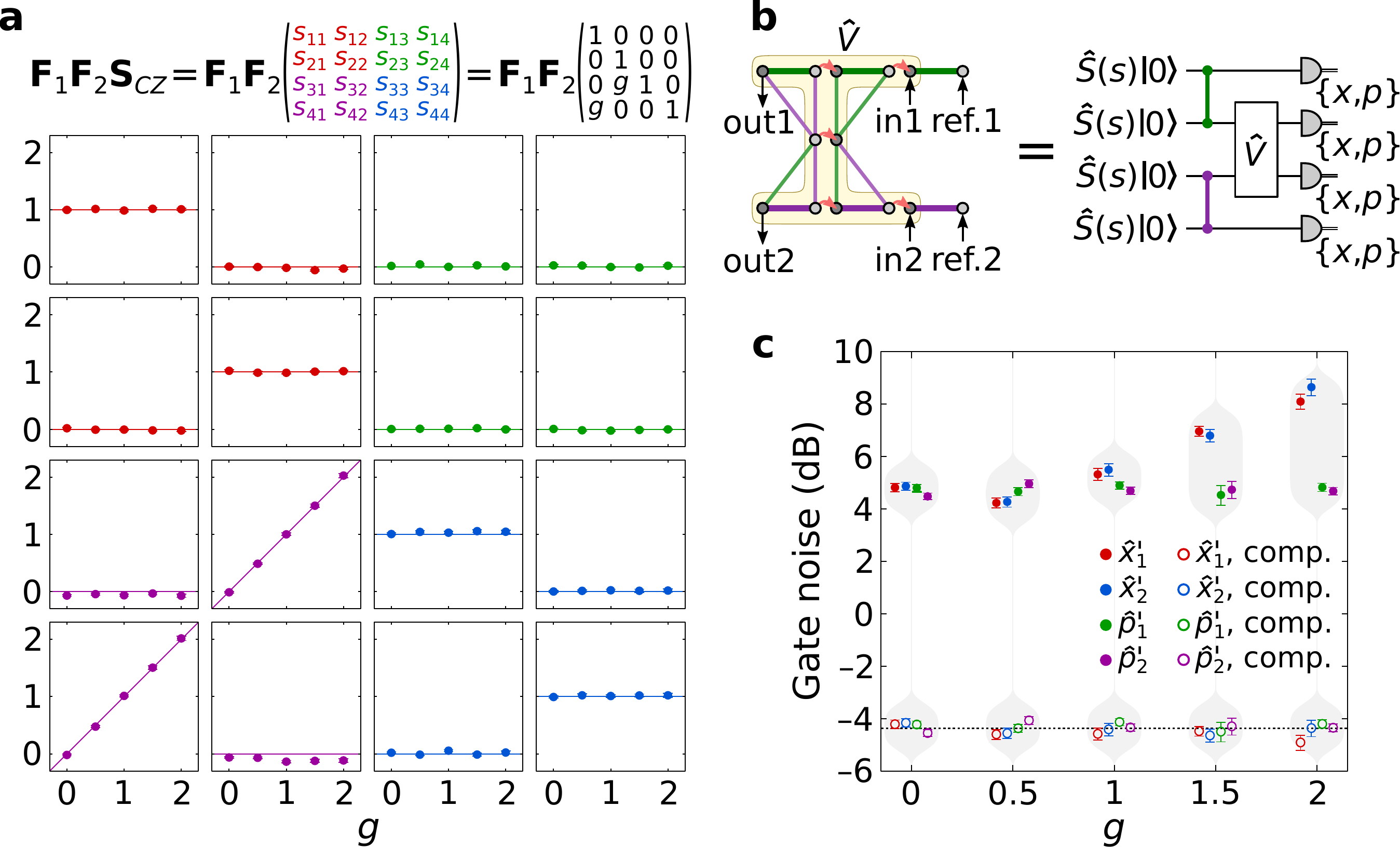}
	\caption{\label{fig:CZ_gate}
	\textbf{Two-mode gate.} \textbf{a}, Symplectic matrix corresponding to \textbf{G} in Eq.~\eqref{eq:operation} for the implemented controlled-Z gate (modified by $\hat{F}\otimes\hat{F}$) as a function of the coupling coefficient $g$, measured by gate tomography (described in SI section 3) with the corresponding circuit shown in \textbf{b}. \textbf{c}, Measured gate noise in the four output quadratures for each implemented controlled-Z gate (solid points), together with gate noise compensated for the effect of the gate noise matrix $\textbf{N}$ in Eq.~\eqref{eq:operation} (hollow points). The values for compensation are given in the SI Fig. S6 for each value of $g$. The compensated gate noise can be compared with the initial squeezing variance for cluster state generation, measured to be $\SI{-4.4}{dB}$ (dashed line). Estimation of error bars are described in Fig.~\ref{fig:single_mode}.	}
\end{figure*}

We characterize the implemented gates with gate tomography by letting the input mode be entangled to a reference mode and measuring the quadrature correlations between the reference and the gate-teleported output modes \cite{asavanant20}. The resulting symplectic matrices are shown in Fig.~\ref{fig:single_mode}a with the corresponding circuit summarized in Fig.~\ref{fig:single_mode}b---for details, see SI section 3.1. The symplectic matrix elements are seen to agree well with the theoretical values, certifying the accuracy of the gates. The same measurements reveal the added gate noise, the other pertinent parameter characterizing the gates' performance. For GKP qubit computation, quadrature gate noise is the most relevant figure of merit, since this is what eventually causes qubit errors. Fault-tolerance thresholds are therefore traditionally presented in terms of quadrature noise, i.e. quadrature squeezing\cite{menicucci14,fukui18,walshe19,larsen20}. The gate noise is shown in Fig.~\ref{fig:single_mode}c. For our computation scheme, we expect a gate noise of $\text{Var}\{\textbf{N}\boldsymbol{\hat{p}_i}\}=4e^{-2r}V_0$ (SI section 2.1) which is four times larger than the initially squeezed state variance of $e^{-2r}V_0$ (where $V_0$ is the vacuum variance). Therefore, by compensating the measured gate noise by the four vacuum units, $1/4\approx\SI{-6}{dB}$, we expect to regain the initially measured squeezing variance of $\SI{4.4}{dB}$. The compensated noise level is illustrated in Fig.~\ref{fig:single_mode}c and is seen to agree well with the expected value except for the squeezing gate where the squeezing level, $e^r=\tan\theta_-/2$, becomes highly sensitive to phase fluctuations in $\theta_-=-\theta_{A,k}+\theta_{B,k}$ for large $|r|$.

To further demonstrate the impact of the cluster state entanglement, the measured gate noise is plotted in Fig.~\ref{fig:single_mode}d as a function of the OPO pump power that controls the squeezing process. For vanishing squeezing (zero pump power) where the cluster state is simply a vacuum state, gate noise of $\SI{6}{dB}$ is measured, corresponding to the classical limit of our scheme. When increasing the OPO pump power, the gate noise reduces below this limit due to the increasing cluster state entanglement. The measured gate noise agrees well with that estimated from the experimental parameters of the setup~\cite{larsen19}. Obviously, for fault-tolerant computation, much lower gate noise is required. Gate noise with potential improvements of the setup are estimated and plotted in Fig.~\ref{fig:single_mode}d as well. It is clear that higher optical efficiencies and larger squeezing bandwidth significantly decrease the gate noise and can bring the system towards fault-tolerant computation. See SI section 4 for a more comprehensive discussion on gate noise.

To complete the universal Gaussian gate set, we implemented a two-mode gate---a modified version of the controlled-Z gate, $(\hat{F}\otimes\hat{F}^j)\hat{C}_Z(g)=(\hat{F}\otimes\hat{F}^j)e^{ig\hat{x}\otimes\hat{x}}$. Here $w$ in $j=(-1)^w$ is the lower wire number of the two wires on which the gate is implemented. The required basis setting and resulting gate noise is derived in SI section 2.2. To shorten the notation, in the following we denote $(\hat{F}\otimes\hat{F}^j)$ simply as $\hat{F}\hat{F}$. Note that the transformation $\hat{F}\hat{F}$ can be easily reversed in subsequent transformations to realize a pure $\hat{C}_Z(g)$ gate~\cite{larsen20}. Together with the implemented single-mode gates, $\{\hat{F}\hat{F}\hat{C}_Z(g),\hat{R}(\theta),\hat{S}(e^r)\}$ constitutes a universal multi-mode Gaussian gate-set \cite{lloyd99}, while $\{\hat{F}\hat{F}\hat{C}_Z(1),\hat{R}(\pi/2),\hat{F}^j\hat{P}(1)\}$ constitutes a multi-mode Clifford gate-set on GKP-encoded qubits \cite{gottesman01}. For gate tomography of the implemented $\hat{F}\hat{F}\hat{C}_Z(g)$-gate, quadrature correlations of the output state and reference states entangled to the input states are measured (see SI section 3.2). The resulting symplectic matrix, together with the corresponding gate tomography circuit, is shown in Figs.~\ref{fig:CZ_gate}a and \ref{fig:CZ_gate}b, and the measured symplectic matrix elements are seen to agree well with the expected values. The gate noise, shown in Fig.~\ref{fig:CZ_gate}c, is larger than for single-mode gates since two-mode gates are implemented in two computation steps and depend on $g$. By compensating for the effect of $\textbf{N}$, we again retrieve the expected squeezing variance of $-\SI{4.4}{dB}$ with good agreement.

\begin{figure*}
	\includegraphics[width=0.77\linewidth]{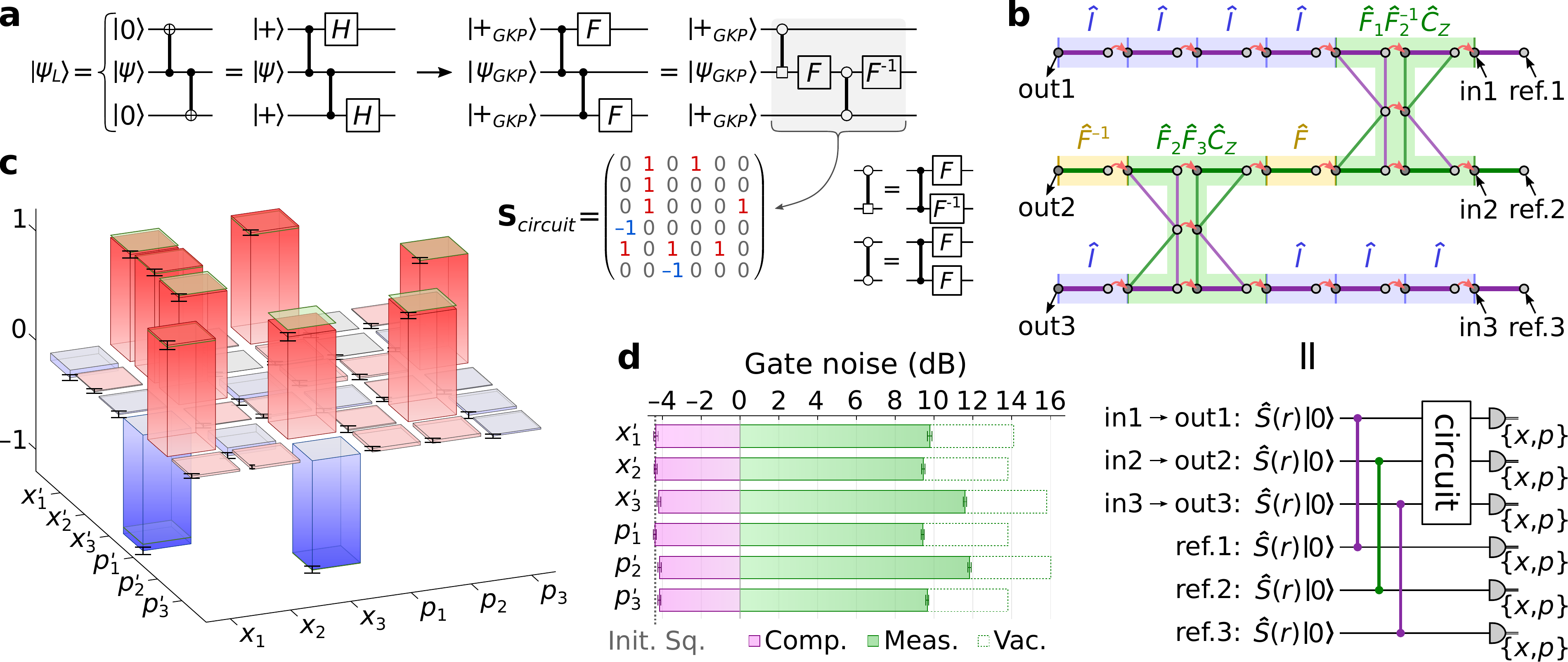}
	\caption{\label{fig:circuit}
	\textbf{Quantum circuit.} \textbf{a}, Circuit encoding a logic qubit in the 3-qubit bit-flip error correction code, rewritten in terms of the CV $\hat{F}\hat{F}\hat{C}_Z(1)$ and $\hat{F}^{\pm1}=\hat{R}(\pm\pi/2)$ gates to take GKP-encoded qubits as input. \textbf{b}, Implementation of the encoding circuit (marked by gray area in \textbf{a}) on three coupled cluster state wires, together with the corresponding circuit for circuit tomography described in SI section 3.3. \textbf{c}, Resulting circuit symplectic matrix estimated from quadrature correlations of input and reference modes in the circuit tomography. Here green pads show the expected values, listed for $\textbf{S}_\text{circuit}$ in \textbf{a}. \textbf{d}, Measured gate noise (green bars), compared to expected gate noise for vacuum in place of the cluster state (dashed bars). Also, gate noise compensated for the combined circuit gate noise matrix \textbf{N} is shown (purple bars), with compensated values given in SI Fig. S7. The compensated gate noise can be compared with the initial squeezing variance for cluster state generation of $\SI{-4.4}{dB}$ (dashed gray line). Estimation of error bars are described in Fig.~\ref{fig:single_mode}.}
\end{figure*}

\subheading{Quantum circuit}
To demonstrate the flexibility of combining gates into a programmable quantum circuit, here we implement as an example a three-mode circuit, which for three physical GKP qubits as input states encodes a logical qubit in the three-qubit bit-flip error correction code (see Fig.~\ref{fig:circuit}a). The implementation of the circuit on the cluster state is illustrated in Fig.~\ref{fig:circuit}b. It includes two $\hat{F}\hat{F}\hat{C}_Z(1)$ gates corresponding to qubit controlled-Z and Hadamard gates, two $\hat{F}^{\pm1}=\hat{R}(\pm\pi/2)$ gates corresponding to qubit Hadamard gates, and eight identity gates which can be thought of as qubit memory and may be unnecessary depending on the surrounding circuit. The details of the implementation and associated gate noise are discussed in SI section 2.3.

To characterize the performance of the implemented circuit, we perform circuit tomography similar to the strategy applied for gate transformations (see SI section 3.3). The resulting gate symplectic matrix, shown in Fig.~\ref{fig:circuit}c, is seen to resemble the desired matrix. In Fig.~\ref{fig:circuit}d, the measured gate noise is shown and compared to the expected gate noise for a cluster state with no entanglement. It is clear that the entanglement of the cluster state leads to a reduction of the gate noise. To verify the measured gate noise values, we back-propagate the combined gate noise through the circuit by compensating for $\textbf{N}$ (as presented in Fig.~S7 of SI) with the result shown in Fig.~\ref{fig:circuit}d, and the estimated values agree well with the initially measured degree of squeezing of $\SI{4.4}{dB}$. Note that the large total circuit noise stems from the accumulation of gate noise associated with multiple concatenated gates and the lack of error correction. To prevent gate noise accumulation as required for fault-tolerant computation, GKP quadrature error correction should be performed as often as possible, preferably in between each implemented gate \cite{larsen20}.
 
\subheading{Outlook}
We have demonstrated the machinery for performing MBQC on our cluster state architecture, which relies on comparatively low-tech photonic technology at room temperature, and illustrated its computational flexibility by combining 12 gates into a programmable quantum circuit. The single- and two-mode gates can be organized in any order on the six input modes of the cluster state, thereby allowing for the implementation of an arbitrary six-mode circuit transformation of, in principle, infinite depth. The demonstrated platform is currently restricted to a six-mode circuit, but due to its inherent deterministic nature, the platform can be efficiently up-scaled to allow for large-scale computation. This can be attained by increasing the bandwidth of the optical squeezing process and complement it with broadband homodyne detectors. Bandwidths of several GHz are possible~\cite{Tasker2020,kashiwazaki20}, so the number of input modes can be increased to several thousands, bringing the platform well into NISQ (noisy intermediate-scale quantum technology) territory~\cite{Preskill2018,su18}. Furthermore, with the platform's telecom compatibility, multiple processing units may straight-forwardly be combined and scaled up without the need of complex quantum transduction. More generally, instead of the all-temporal encoding used here for constructing and scaling the optical cluster state, it is also possible to use spectral~\cite{cai17,chen14} and spatial~\cite{armstrong12} degrees of freedom.

To attain fault-tolerant universal quantum computing on our platform, the gate noise must be significantly decreased and the quantum information must be encoded as qubits, such as GKP qubits. Fault-tolerance is attained by increasing the amount of squeezing of the cluster state to lower the gate noise and by using GKP ancilla states for quantum error-correction to prevent the accumulation of noise. Such error-correction can be done directly in our computation scheme (with homodyne detection and without active feedforward) simply by replacing certain sqeezed states of the 1D cluster with GKP states\cite{walshe20}. The squeezing threshold for fault-tolerance of similar MBQC schemes has been estimated to be in the range of 10--$\SI{17}{dB}$ \cite{walshe19,fukui18,larsen21} which should be compared to the state-of-the-art of squeezing of $\SI{15}{dB}$\cite{vahlbruch16}. GKP states have been generated in vibrational modes of trapped ions~\cite{Fluhmann2019} and in microwave cavity fields~\cite{campagne20} but it remains a challenge to produce them in the optical spectrum although proposals exist~\cite{Pirandola2006,Motes2017,Weigand2018,Eaton2019,su19,tzitrin20}. Once this challenge has been solved, all basic ingredients for fault-tolerant, universal, scalable quantum computing are available.

	\vspace{4mm}
	\noindent\textbf{Acknowledgements:} We acknowledge useful discussion with R. N. Alexander and J. Hastrup. The work was supported by the Danish National Research Foundation through the Center for Macroscopic Quantum States (bigQ, DNRF0142).\\
    
    \noindent\textbf{Author contributions:} M.V.L. and U.L.A. conceived the project. J.S.N., X.G., C.R.B., and M.V.L. built the squeezing sources. M.V.L. developed the theoretical background, designed the experiment and build the setup. M.V.L. performed the experiment and data analysis. The project was supervised by U.L.A. and J.S.N. The manuscript was written by U.L.A., M.V.L. and J.S.N. with feedback from all authors.\\
    
    \noindent\textbf{Data availability:} Datasets generated during this study, and code for data analysis, are available at https://doi.org/10.11583/DTU.14673570\\

\end{document}


\newpage
	\title{Supplementary Information for\\\vspace{0.2cm}Deterministic multi-mode gates on a scalable\\photonic quantum computing platform}
	\author{Mikkel V. Larsen, Xueshi Guo, Casper R. Breum,\\Jonas S. Neergaard-Nielsen and Ulrik L. Andersen}
	\date{October 27, 2020}
	\maketitle
	\vspace{2cm}
	\tableofcontents	

	\newpage

\section{Experimental setup}\label{sec:setup}
\begin{figure}[b!]
	\centering
	\includegraphics[width=0.85\textwidth]{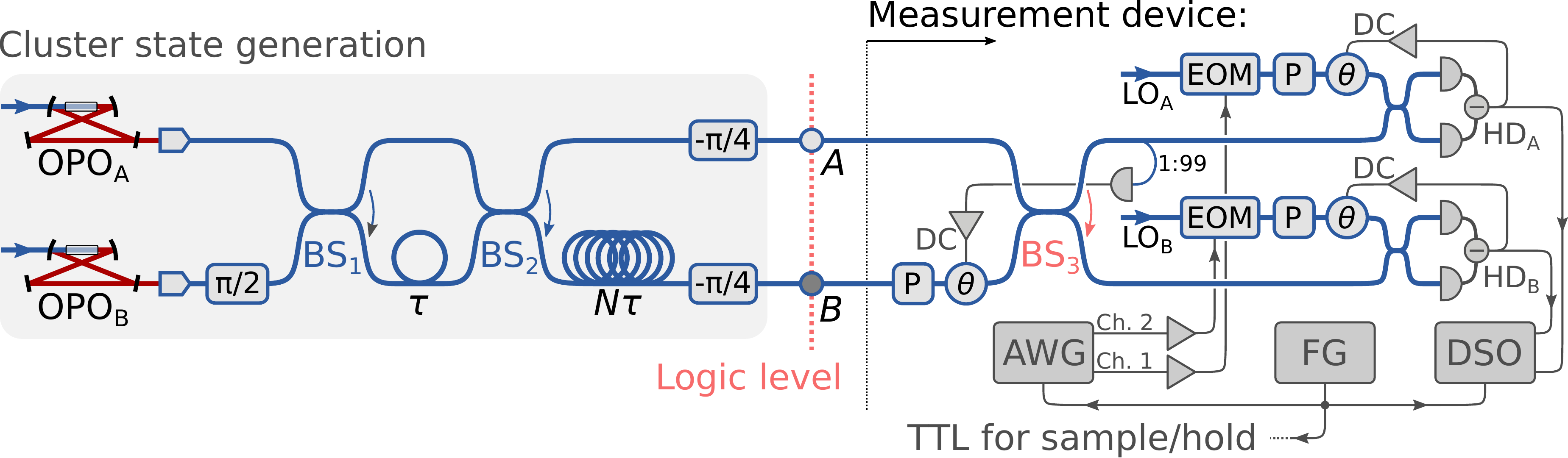}
	\caption{Schematics of the experimental setup showing the details of the measurement device for gate implementation by projective measurements, while details on the setup for cluster state generation can be found in the supplementary information of \cite{larsen19}. Red and blue represents free-space and single-mode fiber optics respectively. In the measurement device P indicates polarization control while $\theta$ indicates active phase control using a fiber stretcher \cite{larsen19b}. The required basis setting sequence in the homodyne detectors (HD) to implement a desired gate is controlled using electro-optical modulators (EOM) in the local oscillators (LO). The EOM is driven using an amplified two-channel arbitrary waveform generator (AWG), and the measured quadrature time traces are acquired on a digital storage oscilloscope  (DSO). The AWG, scope and sample/hold locking scheme is triggered from a function generator (FG).}
	\label{fig:setup}
\end{figure}

The experimental setup in the main text Fig.~1b is shown here in Fig.~\ref{fig:setup} with more details of the measurement device including control of the basis settings for different temporal modes. The cluster state generation scheme is similar to the scheme presented in Ref. \cite{larsen19}, where a detailed description of the cluster state generation setup can be found in the corresponding supplementary information. To summarize: Two-mode squeezing in spatial mode $A$ and $B$ is generated by interfering two single mode squeezed states in a beam splitter denoted $\text{BS}_1$. By delaying one mode of the two-mode squeezed states by $\tau$ and subsequently interfering the two modes ($A$ and $B$) at beam-splitter $\text{BS}_2$, a one-dimensional (1D) cluster state, namely a dual-rail wire \cite{menicucci11b,yokoyama13} with temporal mode duration of $\tau$, is formed. The 1D cluster state is coiled up by a long delay of $N\tau$ to form a cylinder with $N$ temporal modes in the circumference. Locally on the cylinder surface the coiled-up cluster state can be pictured as a 2-dimensional (2D) cluster state of parallel 1D cluster states. In \cite{larsen19} the coiled-up 1D cluster state is then interfered with itself by beam-splitter $\text{BS}_3$ to form a double bilayer square lattice \cite{larsen19,larsen20}. However, for cluster state computation, here we will consider the coiled-up cluster state just before $\text{BS}_3$, while $\text{BS}_3$ is part of a joint measurement device for implementing quantum gates by projective measurements. As such, in the experimental setup, the modes in computation are located just before $\text{BS}_3$, and is marked as the \textit{logic level} in Fig.~\ref{fig:setup}. A computation scheme in this logic level is presented in \cite{larsen20}, while here we use a slightly modified scheme as presented in section \ref{sec2}.

In Ref. \cite{larsen19}, the generated coiled up 1D cluster state is in practice a $\mathcal{H}$-graph state \cite{menicucci11a}. Here, we rotate the phases of every mode in the logic level by $-\pi/4$ to transform this $\mathcal{H}$-graph state into a cluster state with real edges. The  $-\pi/4$ phase-rotations on both spatial mode $A$ and $B$ commutes with $\text{BS}_3$, and thus for simplicity, in the experimental realization we apply the $-\pi/4$ phase-rotations directly onto the the local oscillators, $\text{LO}_A$ and $\text{LO}_B$, of the two homodyne detectors $\text{HD}_A$ and $\text{HD}_B$. Thus, in the following, the generated state is considered as a cluster state, and not a $\mathcal{H}$-graph state.

In the measurement device, the phase relation of the input modes of $\text{BS}_3$ is actively locked by tapping off and detecting about $1\%$ of the power in spatial mode $A$, and the DC measurement value is fed back via a PID to a fiber phase controller---a so-called DC-lock. The LO phases of the homodyne detectors are phase-locked in similar fashion using DC-locks, thereby enabling stable measurements of the $\p$- and $\x$-quadrature of the spatial modes $A$ and $B$. Finally, using an electro-optical phase modulator (EOM) in each local oscillator, the homodyne detector bases can be  individually and dynamically controlled in each spatial mode for different temporal modes. To prevent technical noise from the probe beams used for phase-locking of $\text{BS}_\text{1--3}$ and the homodyne detectors, a sample/hold locking scheme is used, where a probe beam is chopped on and off as described in Ref. \cite{larsen19}. All phase locks are activated when the probe is on, while when the probe is off, the phase-lock feedback is kept constant while acquiring data with basis settings controlled for each mode by the EOMs to implement a desired gate. Quadrature time traces from the homodyne detectors are acquired on a digital storage oscilloscope (DSO), and for each temporal mode $k$ the measurement outcome is extracted using the corresponding temporal mode function
\begin{equation}\label{eq:modeFunction}
	f_k(t)=\begin{cases}
		\mathcal{N}(t-k\tau)e^{-\kappa^2t^2} & ,\;|t-k\tau|<\tau/2\\
		0 & ,\;\text{otherwise}
	\end{cases}\;,
\end{equation}
where $\mathcal{N}$ is a normalization factor of units $\SI{}{s^{-1}}$. Here, $\kappa$ is optimized to be $2\pi\times\SI{2.0}{MHz}$ to minimize the gate noise. In the data acquisition we measure $228$ temporal modes, consecutively, corresponding to $19$ turnarounds of the cylindrical cluster state for $N=12$. As shown in section \ref{sec3}, this allows us to implemented and characterize multiple gates at once in parallel.

The experimental setup is operated at the telecom wavelength of $\SI{1550}{nm}$ to minimize propagation losses in optical fibers (blue lines in Fig.~\ref{fig:setup}). For the short delay line a $\SI{50}{m}$ fiber is used leading to $\tau\approx\SI{250}{ns}$ temporal mode duration, while for the long delay line $\SI{600}{m}$ is used leading to $N=12$. For phase locks, fiber stretchers of negligible optical losses are used \cite{larsen19b}. The EOMs for setting the homodyne detection bases are of model MPZ-LN-10 from iXblue with $\SI{10}{GHz}$ bandwidth. To control the EOMs, a two channel arbitrary waveform generator (AWG) of model M4i.6631-x8 from Spectrum Instruments with $\SI{1.25}{GS/s}$ sampling rate and $\SI{400}{MHz}$ bandwidth is used to generate the required waveforms to measure in a basis setting sequence implementing a desired gate. The waveform signals from the AWG driving the EOMs are amplified using THS3491 operational amplifiers (op-amps) from Texas Instrument with a $\SI{8000}{V/\micro s}$ slew rate. To compensate for electrical responses in the AWG, op-amps and EOMs, before the experiment is carried out the AWG waveforms are optimized by inserting each EOM in a Mach–Zehnder interferometer, and feeding back the resulting applied phase shift from the EOMs to the AWG in order to update the waveform targeting a desired phase shift sequence. As a result, we are able to switch the homodyne detection basis from $-\pi/2$ to $\pi/2$ (both stable within $1\%$ of the value) within $\SI{8}{ns}$, followed by a constant phase of the desired value after switching.

\section{Computation scheme}\label{sec2}
In the experimental setup shown in the main text Fig.~1b, and explained in detail in section \ref{sec:setup}, a dual-rail 1D cluster state \cite{menicucci11b,yokoyama13} is generated and coiled up into a cylinder with $N$ temporal modes in the circumference to form a cluster state with a local 2D topology. The 1D cluster state can be used for computation along the cluster state \cite{alexander14,asavanant20}, but here, with the cluster state coiled up, the goal is instead to perform computation with information flowing along the cylinder, i.e. across the 1D cluster state. The scheme for doing so is explained in the following, and the required basis settings for implementing single-mode and two-mode gates are described in section \ref{sec2:single} and \ref{sec2:two}, while in section \ref{sec2:circuit} we combine single- and two-mode gates to implement a circuit.

\begin{figure}
	\centering
	\includegraphics[width=0.78\textwidth]{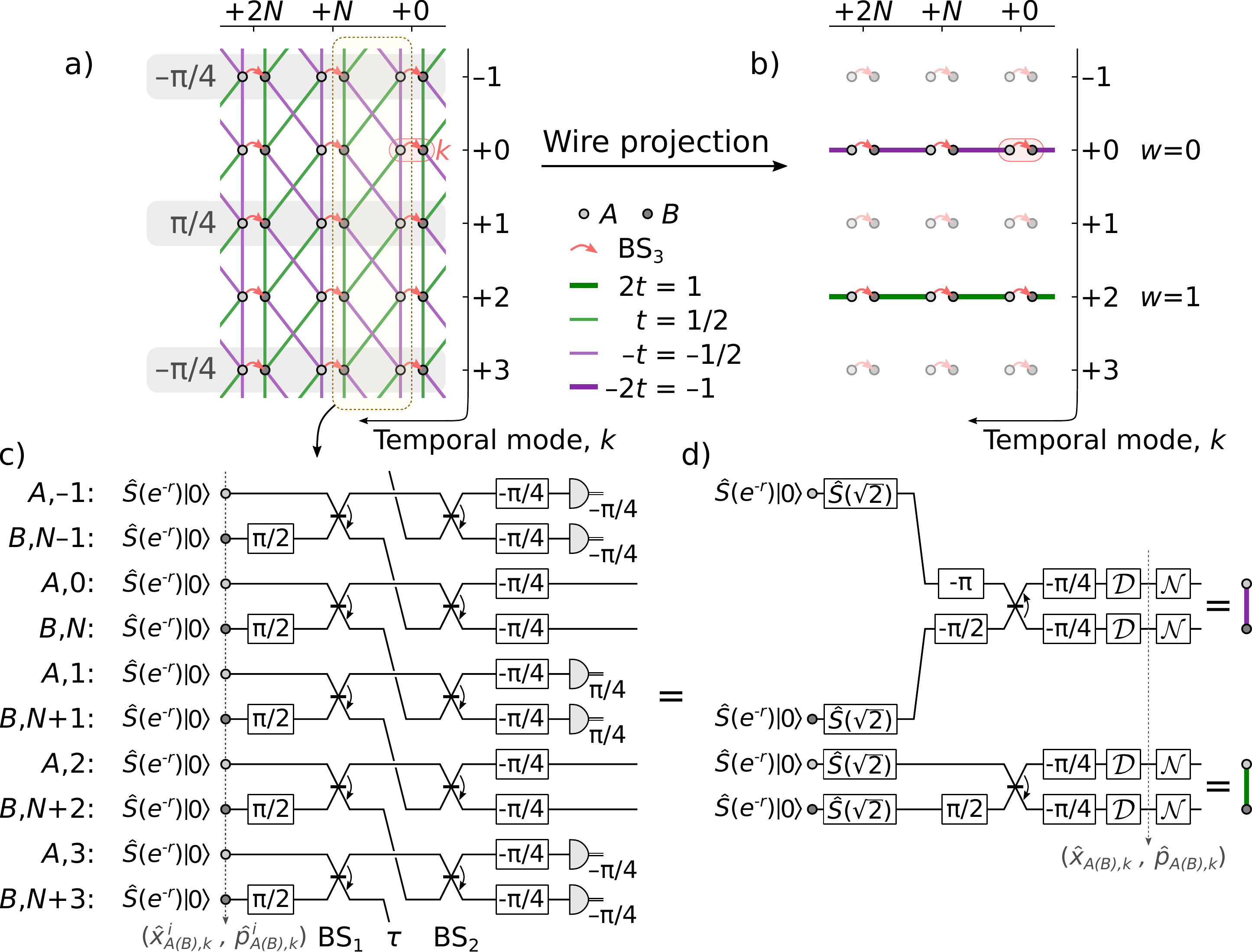}
	\caption{(a) Section of the coiled-up dual-rail 1D cluster state in the logic level of the main text Fig.~1b, forming a local 2D cluster state of parallel vertical 1D cluster states. Here the bright and dark modes represent spatial mode $A$ and $B$, respectively, while temporal mode $k=0$ is marked as a red area. Modes in the gray shaded area are control modes, which are measured in basis $\x(\pm\pi/4)$ to project the 2D cluster state into wires. The red arrow in each temporal mode indicates the beam-splitter of the measurement device, $\text{BS}_3$, and can for the control modes be compensated for as shown in Eq.~\eqref{eq:compBS3}. (b) After measuring control modes, the cluster state is projected into wires with wire number $w=(k\mod  N)/2$. (c) Shows the 1D cluster state generation and wire projection, while (d) shows the corresponding simplified circuit after measurements. Here $\mathcal{D}$ and $\mathcal{N}$ corresponds to displacements and noise as described in Eq.~(\ref{eq:D}) and (\ref{eq:Ne},\ref{eq:No}).}
	\label{fig:projection}
\end{figure}

In the language of graphical calculus of Gaussian states \cite{menicucci11a}, the generated 1D cluster state has edge weights of $\pm t=\pm\tanh(2r)/2$ and self-loops of $i\sech(2r)$ \cite{menicucci11b}, where $r$ is the squeezing parameter of the initial momentum squeezed states. This is an approximate cluster state with the variance of the nullifiers,
\begin{equation}\begin{aligned}\label{eq:nullifier}
	\hat{n}_{A,i}&=\p_{A,i}-t\left(-\hat{x}_{A,i-1}-\hat{x}_{A,i+1}-\hat{x}_{B,i+N-1}+\hat{x}_{B,i+N+1}\right)\\
	\hat{n}_{B,i}&=\p_{B,i}-t\left(\hat{x}_{A,i-N-1}-\hat{x}_{A,i-N+1}+\hat{x}_{B,i-1}+\hat{x}_{B,i+1}\right)\;,
\end{aligned}\end{equation}
vanishing in the limit of infinite squeezing:
\begin{equation}\label{eq:NullVar_approx}
	\text{Var}\{\hat{n}_{A,i}\}=\text{Var}\{\hat{n}_{B,i}\}=V_0\sech(2r)\rightarrow0\quad,\quad \text{for }r\rightarrow\infty\;,
\end{equation}
where $V_0=1/2$ is the variance of vacuum for $\hbar=1$, and the subscript indicates the mode numbering in Fig.~\ref{fig:projection}. Since the edge weight, $t$, depends on the squeezing parameter, $r$, the required basis setting for implementing a desired quantum gate will in general depend on $r$. Such scheme is described in detail in Ref. \cite{larsen20} for the cluster state considered here, namely the double bilayer square lattice (DBSL). However, in practice, it is inconvenient to have the basis settings depend on $r$, as the exact initial squeezing most often is unknown at the point the experiment is carried out and may vary slightly from time to time. Instead, here we redefine the generated state to have the same graph but with edge weights $\pm t=\pm 1/2$, i.e. we drop the squeezing dependent $\tanh(2r)$ in the edge weights. The resulting 1D graph state is by the definition in Ref. \cite{vanloock07} a \textit{cluster-type state}\footnote{A cluster-type state, with graph vertices $a\in G$ and connected nodes $b\in N_a$ to $a$, is defined in \cite{vanloock07} as a multi-mode Gaussian state where the variance of $\p_a-\sum_{b\in N_a}t_b\x_b$ (here the non-zero variable $t_b$ is added by us to generalize the cluster-type states to have variable edge weights) vanishes in the limit of infinite squeezing. Thus, cluster-type states are a more general group of states allowing cluster state approximations not covered by graphical calculus for Gaussian pure states \cite{menicucci11a}.} with the variance of the nullifiers in Eq.~\eqref{eq:nullifier} equal
\begin{equation}\label{eq:NullVar_Type}
	\text{Var}\{\hat{n}_{A,i}\}=\text{Var}\{\hat{n}_{B,i}\}=2V_0e^{-2r}\rightarrow0\quad,\quad \text{for }r\rightarrow\infty\;,
\end{equation}
which as well vanish in the limit of infinite squeezing. Note that at high squeezing levels where $\sech(2r)\approx2e^{-2r}$, the nullifier variances for the approximate cluster state in Eq.~\eqref{eq:NullVar_approx} and the cluster-type state in Eq.~\eqref{eq:NullVar_Type} are equal, while for vanishing squeezing, \eqref{eq:NullVar_approx} approaches the vacuum variance, $V_0$, while \eqref{eq:NullVar_Type} approaches $2V_0$. Thus, for finite squeezing, the redefined cluster-type state with $t=1/2$ is more noisy than the more traditional approximate cluster state with $t=\tanh(2r)/2$. Fortunately, we find that the noise produced by gates implemented on the generated graph state considered as a cluster-type state is at the same level or, in some cases, even lower than if we use the graph state as an approximate cluster state. In the following, unless it may lead to confusion, we will use the term `cluster state' for both cluster-type states and approximate cluster states.

The rules for graphical calculus of Ref. \cite{menicucci11a} applies to approximate cluster states and are therefore, to our knowledge, not necessarily valid for the more general cluster-type states. As a result, here we will derive the necessary graph transformation. A section of the coiled-up 1D cluster state is shown in Fig.~\ref{fig:projection}a. To perform computation, we need to project the cluster state into \textit{wires} along the cylinder, on which gates can be implemented \cite{larsen20}. To do so, \textit{control modes} (in the grey shaded areas of Fig.~\ref{fig:projection}a) are measured in alternating bases of $(-1)^{(k-1)/2}\pi/4$ where odd $k$ is the mode number of the temporal control modes in Fig.~\ref{fig:projection}a, i.e. measuring $\x(\pm\pi/4)=(\x\pm\p)/\sqrt{2}$ of control modes by homodyne detection. From standard graphical calculus of Ref. \cite{menicucci11a}, we expect this measurement to form new edges of weight $\pm 2t$ as shown in Fig.~\ref{fig:projection}b. To see that this is indeed the case for the cluster-type state with $t=1/2$, we consider the 1D cluster state generation scheme in Fig.~\ref{fig:projection}c: After measuring the control modes, the generation scheme can be simplified to separate the generation of two-mode entangled states as shown in Fig.~\ref{fig:projection}d (to derive this, we have used the method of calculating quadrature transformations outlined in appendix A of Ref. \cite{larsen20}). Here $\mathcal{D}$ symbolises a displacement in phase-space, $\hat{D}(\alpha)$, by $\text{Re}\lbrace\alpha\rbrace$ and $\text{Im}\lbrace\alpha\rbrace$ in $\x$- and $\p$-quadrature, respectively, depending on the control mode measurement outcomes as
\begin{equation}\begin{aligned}\label{eq:D}
	\D_{A,k}(\alpha)\;,\;\alpha&=\frac{1}{2\sqrt{2}}\Big[(j-i)m_{A,k-1}+(-j-i)m_{A,k+1}+(j-i)m_{B,k+N-1}+(j+i)m_{B,k+N+1}\Big]\;,\\
	\D_{B,k+N}(\beta)\;,\;\beta&=\frac{1}{2\sqrt{2}}\Big[(-j+i)m_{A,k-1}+(-j-i)m_{A,k+1}+(-j+i)m_{B,k+N-1}+(j+i)m_{B,k+N+1}\Big]\;,
\end{aligned}\end{equation}
where $j=(-1)^w$ with $w=(k\mod N)/2$ being the wire number, while $i$ is the imaginary unit, $i^2=-1$. Since the measurement outcomes, $m_{A(B),k}$, are known, this displacement can be compensated for by feeding the measurement results forward to displacement operations displacing the wire modes back. Or, we can simply keep track of the displacements and compensate for them in the final measurement outcomes since all operations implemented here are Gaussian operations \cite{menicucci06,gu09}. $\mathcal{N}$ symbolises a quadrature symmetric noise operator, and originate from finite squeezing in the 1D cluster state generation. In the Heisenberg picture, this is represented by adding initial finite squeezed momentum quadratures, $\p_{A(B),k}^i$, to the the wire cluster state quadratures $\x_{A(B),k}$ and $\p_{A(B),k}$. Here $(\x_{A(B),k}^i,\p_{A(B),k}^i)$ and $(\x_{A(B),k},\p_{A(B),k})$ are marked on Fig.~\ref{fig:projection}c,d. The resulting quadrature evolution of $\mathcal{N}$ on the wire mode quadratures is
\begin{equation}\label{eq:Ne}
	\begin{pmatrix}
	\x_{A,k}\\ \x_{B,k+N} \\ \p_{A,k}\\ \p_{B,k+N} 
	\end{pmatrix}\xrightarrow{\mathcal{N}}
	\begin{pmatrix}
	\x_{A,k}\\ \x_{B,k+N} \\ \p_{A,k}\\ \p_{B,k+N}
	\end{pmatrix}+\frac{1}{2\sqrt{2}}
	\begin{pmatrix}
	-1 & -1\\ -1 & 1 \\ 1 & -1\\ 1 & 1
	\end{pmatrix}
	\begin{pmatrix}
		\p_{A,k}^i\\ \p_{B,k+N}^i
	\end{pmatrix}
\end{equation}
for even wires ($w=(k\mod N)/2=\text{even}$), and
\begin{equation}\label{eq:No}
	\begin{pmatrix}
	\x_{A,k}\\ \x_{B,k+N} \\ \p_{A,k}\\ \p_{B,k+N}
	\end{pmatrix}\xrightarrow{\mathcal{N}}
	\begin{pmatrix}
	\x_{A,k}\\ \x_{B,k+N} \\ \p_{A,k}\\ \p_{B,k+N} 
	\end{pmatrix}+\frac{1}{2\sqrt{2}}
	\begin{pmatrix}
	1 & -1\\ -1 & -1 \\ -1 & -1\\ 1 & -1
	\end{pmatrix}
	\begin{pmatrix}
		\p_{A,k-1}^i\\ \p_{B,k+N+1}^i
	\end{pmatrix}
\end{equation}
for odd wires ($w=(k\mod N)/2=\text{odd}$). Since the initial momentum squeezed quadratures, $\p_{A(B),k}^i$, squeezed by $e^{-r}$, has a Gaussian quadrature distribution with variance $\text{Var}\{\p_{A(B),k}^i\}=e^{-2r}V_0$ and mean $\braket{\p_{A(B),k}^i}=0$, the $\mathcal{N}$-operation adds noise to the wire mode quadratures depending on the initial momentum squeezing, which vanish in the limit of infinite squeezing. From Eq.~\eqref{eq:Ne} and \eqref{eq:No} the noise added by $\mathcal{N}$ is seen to be correlated in the quadratures between two connected wire modes. This is exactly in such a way, that $\mathcal{N}$ can be brought to the left side of the beam-splitter and phase shifts in Fig.~\ref{fig:projection}d, in which case $\mathcal{N}$ adds uncorrelated noise in the two wire-modes before the beam-splitter. Fortunately, for the same reason, the quadrature noise added by $\mathcal{N}$ cancels out in the wire cluster state nullifiers and when implementing gates, as we will see later. Finally, note the $\sqrt{2}$ anti-squeezing on the initial momentum squeezed states in Fig.~\ref{fig:projection}d. This anti-squeezing leads to a degradation of the ``effective'' initial momentum squeezed variance which becomes $2e^{-2r}V_0$ instead of $e^{-2r}V_0$, and is the cost of projecting the cluster state into another cluster state useful for computation \cite{larsen20}.

Similar to the 1D cluster state with edge weights $\pm t=\pm1/2$, the projected two-mode entangled states in Fig.~\ref{fig:projection}b,d are cluster-type states with edge weights $\pm2t=\pm1$: After compensating for the displacements $\mathcal{D}$, the nullifiers are
\begin{equation*}\begin{aligned}
	\hat{n}_{A,k}&=\p_{A,k}-(-1)\x_{B,k+N}=-\sqrt{2}\p_{A,k-1}^i-\sqrt{2}\p_{B,k+N+1}^i\\
	\hat{n}_{B,k+N}&=\p_{B,k+N}-(-1)\x_{A,k}=\sqrt{2}\p_{A,k-1}^i-\sqrt{2}\p_{B,k+N+1}^i\\
\end{aligned}\end{equation*}
for even wires with $-2t=-1$ edge weight, and
\begin{equation*}\begin{aligned}
	\hat{n}_{A,k}&=\p_{A,k}-\x_{B,k+N}=-\sqrt{2}\p_{A,k}^i+\sqrt{2}\p_{B,k+N}^i\\
	\hat{n}_{B,k+N}&=\p_{B,k+N}-(-1)\x_{A,k}=\sqrt{2}\p_{A,k}^i+\sqrt{2}\p_{B,k+N}^i\\
\end{aligned}\end{equation*}
for odd wires with $2t=1$ edge weight, such that the variances of all nullifiers become
\begin{equation*}
	\text{Var}\{\hat{n}_{A(B),k}\}=4V_0e^{-2r}\rightarrow 0 \text{ for }r\rightarrow\infty\;,
\end{equation*}
and thus vanish in the limit of infinite squeezing as required for cluster-type states. Note that the quadratures added by $\mathcal{N}$ in Eq.~(\ref{eq:Ne},\ref{eq:No}) are not present as they cancel out due to their (anti-)correlations between the connected wire modes in (even)odd wires.

During the projection of the wires in Fig.~\ref{fig:projection}c,d we ignored the beam splitter of the measurement device, $\text{BS}_3$, marked by red arrows in Fig.~\ref{fig:projection}a,b (here we define a balanced beam-splitter operation as $\hat{B}=e^{-i\pi(\x\otimes\p-\p\otimes\x)/4}$ similarly as in Ref. \cite{larsen20,larsen19}, and its direction-dependency is marked below by an arrow pointing from the first to the second mode of the tensor products). This is possible when the two spatial modes $A$ and $B$ of the same temporal mode, $k$, are measured in the same basis, as is the case for the control modes, since
\begin{equation}\label{eq:compBS3}
	\includegraphics[width=0.66\textwidth]{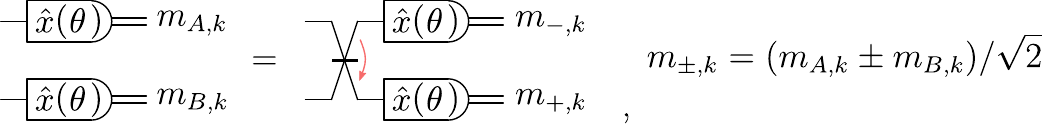}\;.
\end{equation}
Thus, the hypothetical measurement outcomes of the control modes before $\text{BS}_3$, $m_{A(B),k}$, are extracted from the homodyne measurement outcomes of the measurement device after $\text{BS}_3$ as $m_{A,k}=(m_{+,k} +m_{-,k})/\sqrt{2}$ and $m_{B,k}=(m_{+,k}-m_{-,k})/\sqrt{2}$. Note that, since we consider the computation from the logic-level point of view, the notation used here is opposite to that used in the language of macro-nodes in \cite{alexander14,alexander16a,alexander16b}. Here `$A$' and `$B$' refer to spatial mode $A$ and $B$ \textit{before} $\text{BS}_3$, while `$-$' and `$+$' refer to spatial mode $A$ and $B$ \textit{after} $\text{BS}_3$, respectively.

The wire modes in Fig.~\ref{fig:projection}b, projected into two-mode cluster states, are now suitable for the implementation of gates. With $N=12$ temporal modes in the circumference of the cluster state cylinder generated in this work, the coiled-up 1D cluster state is projected into $N/2=6$ wires, numbered as $w\in\lbrace0,1,2,3,4,5\rbrace$ with edge  weight $-(-1)^w$. As a result, the generated cluster state may hold $6$ modes in computation, while this can be scaled up simply by increasing $N$ corresponding to the ratio between the long and short delay lines in the experimental setup.

\subsection{Single mode gates}\label{sec2:single}
\begin{figure}
	\centering
	\includegraphics[width=0.55\textwidth]{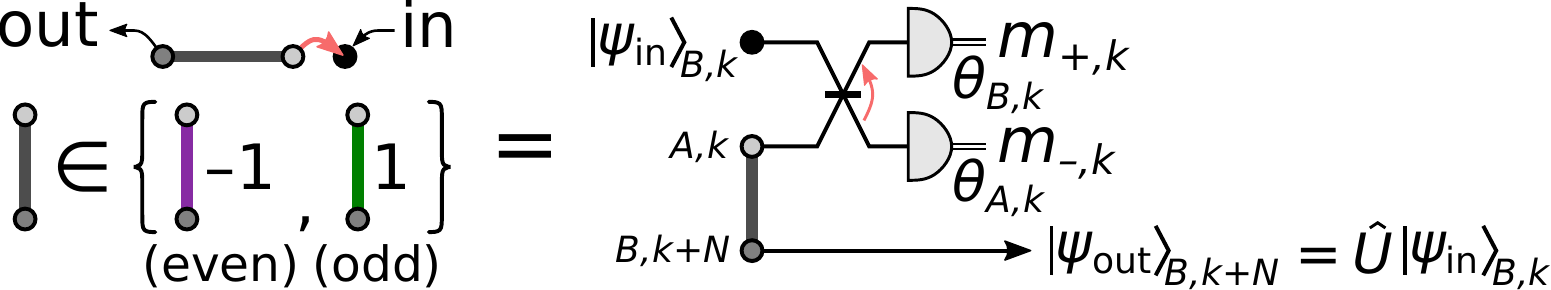}
	\caption{A single computation step from mode $B,k$ to $B,k+N$ on an even (purple with $-1$ edge weight) or odd (green with $1$ edge weight) projected wire in Fig.~\ref{fig:projection}, leading to the implemented operation with the quadrature transformation shown in Eq.~\eqref{eq:U}. The black node represents the input mode, which may be switched into mode $B,k$ using an optical switch in the logic level at spatial mode $B$, or may, as most often, be the output of the previous computation step. The red arrow indicates the beam-splitter $\text{BS}_3$ of the measurement device.}
	\label{fig:single_mode}
\end{figure}
Single-mode gates are implemented on each wire, and a single computation step is shown in Fig.~\ref{fig:single_mode}. The input mode may be switched into the spatial mode $B$ at the logic level using an optical switch, or is, most often, the output mode of the previous computation step. With a joint measurement of the input mode and one mode of the two-mode wire cluster state, using the measurement device consisting of $\text{BS}_3$ and two homodyne detectors, the input mode is teleported to the second mode of the wire cluster state. Depending on the basis setting the input mode undergoes a Gaussian operation, $\hat{U}$, while teleported:
\begin{equation*}
	\ket{\psi_\text{out}}_{B,k+N}=\hat{U}\ket{\psi_\text{out}}_{B,k}\;.
\end{equation*}
Here, $\hat{U}$ consists partly of a desired gate, $\hat{G}$, a displacement by-product, and a noise operation due to finite squeezing. In the Heisenberg picture, $\hat{U}$ transforms the quadratures of the input mode $B,k$ to the output mode $B,k+N$ as
\begin{equation}\label{eq:U}
	\begin{pmatrix}
		\x'_{B,k+N}\\ \p'_{B,k+N}
	\end{pmatrix}=
	\textbf{G}\begin{pmatrix}
		\x_{B,k}\\ \p_{B,k}
	\end{pmatrix}+
	\textbf{N}\begin{pmatrix}
		\hat{p}_{A,k}^{ei}\\ \hat{p}_{B,k+N}^{ei}
	\end{pmatrix}
	+\textbf{D}\begin{pmatrix}
		\tilde{m}_{-,k} \\ \tilde{m}_{+,k}
	\end{pmatrix}\;,
\end{equation}
where the prime denotes the output mode quadratures after gate implementation. In the following we discuss each of the three terms on the right-hand-side. \textbf{G} is the symplectic matrix corresponding to the implemented gate and can be derived to be
\begin{equation}\label{eq:G}
	\hat{G}=(-1)^w\R\left(\frac{\theta_{+,k}}{2}\right)\Sq\left(\tan\frac{\theta_{-,k}}{2}\right)\R\left(\frac{\theta_{+,k}}{2}\right)\quad,\quad\theta_{\pm,k}=\pm\theta_{A,k}+\theta_{B,k}\;,
\end{equation}
where $\hat{R}(\theta)=e^{-i\theta(\x^2+\p^2)/2}$ and $\hat{S}(s)=e^{i\ln(s)(\x\p+\p\x)/2}$ are the rotation and squeezing operations, respectively, with the squeezing parameter $r=\ln(s)$ (leading to squeezing in the $\x$-quadrature for positive $r$). Note, here $\theta_{A,k}$ and $\theta_{B,k}$ are basis settings of the homodyne detectors in spatial mode $A$ and $B$ after $\text{BS}_3$ and should not be confused with the hypothetical measurement outcomes $m_{A,k}$ and $m_{B,k}$ before $\text{BS}_3$. The subscripts `$+$' and `$-$' of $\theta_{\pm,k}$ simply notes the addition and difference of $\theta_{A,k}$ and $\theta_{B,k}$. The term $\textbf{N}(\hat{p}_{A,k}^{ei},\hat{p}_{B,k+N}^{ei})^T$ represents the gate noise due to finite squeezing in the cluster state generation. Here $\hat{p}_{A,k}^{ei}$ and $\hat{p}_{B,k+N}^{ei}$ are the ``effective'' initial momentum quadratures for the two-mode wire cluster state including the $\sqrt{2}$ anti-squeezing contribution from the wire projection shown in Fig.~\ref{fig:projection}d, and so, when including the wire projection, the second term in Eq.~\eqref{eq:U} is
\begin{equation}\label{eq:N}
	\textbf{N}\begin{pmatrix}
		\hat{p}^{ei}_{A,k}\\ \hat{p}^{ei}_{B,k+N}
	\end{pmatrix} =
	\begin{cases}
		\sqrt{2}
		\begin{pmatrix}
			-1 & -1\\ 1 & -1
		\end{pmatrix}
		\begin{pmatrix}
			\p_{A,k-1}^i\\ \p_{B,k+N+1}^i
		\end{pmatrix} &\text{, for } w=\text{even}\vspace{2mm}\\
		\sqrt{2}
		\begin{pmatrix}
			-1 & 1\\ 1 & 1
		\end{pmatrix}
		\begin{pmatrix}
			\p_{A,k}^i\\ \p_{B,k+N}^i
		\end{pmatrix} &\text{, for } w=\text{odd}\;.
	\end{cases}
\end{equation}
Note, as mentioned previously, the noise added by $\mathcal{N}$ does not appear in the implementation of gates as the added noise is correlated between two connected wire-modes. As a result, with $\p_{A(B),k}^i$ being squeezed by $e^{-r}$, symmetric quadrature noise of variance $4V_0e^{-2r}$ is added as gate noise in each computation step, independent on the basis setting and wire number. Here, $\sum_iN_{qi}^2=4$, with $N_{qi}$ being the elements of the gate noise matrix $\textbf{N}$ and $q=1,2$ for single-mode gates, are \textit{gate noise factors}. Assuming identical squeezing in all cluster state modes, the gate noise variances of each quadrature can in general be written as the initial momentum squeezing variance scaled by the gate noise factors of each corresponding quadrature. Finally, $\textbf{D}(\tilde{m}_{+,k},\tilde{m}_{-,k})^T$ is a displacement by-product depending on the measurement outcomes. Here, the tilde indicates that the displacements in the wire projection $\mathcal{D}$ in Fig.~\ref{fig:projection} are included. Writing this out, the third term in Eq.~\eqref{eq:U} is
\begin{equation}\label{eq:by_product_D}
	\textbf{D}\begin{pmatrix}
		\tilde{m}_{-,k} \\ \tilde{m}_{+,k}
	\end{pmatrix}=
	\frac{j\sqrt{2}}{\sin\theta_{-,k}}\begin{pmatrix}
		-\cos\theta_{A,k} &\cos\theta_{B,k}\\
		\sin\theta_{A,k} & -\sin\theta_{B,k}
	\end{pmatrix}\begin{pmatrix}
		m_{+,k}\\m_{-,k}
	\end{pmatrix}+\frac{1}{\sqrt{2}}\begin{pmatrix}
		-j & -j & -j & j\\
		1 & -1 & 1 & 1
	\end{pmatrix}\begin{pmatrix}
		m_{A,k-1}\\m_{A,k+1}\\m_{B,k+N-1}\\m_{B,k+N+1}
	\end{pmatrix}
\end{equation}
where $j=(-1)^w$. Here, the first term is the displacement by-product from the gate implemented by teleportation, while the second term includes displacements from the wire projection. Note that $m_{\pm,k}$ are the measurement outcomes in Fig.~\ref{fig:single_mode} after $\text{BS}_3$, while $m_{A(B),k}$ are measurement outcomes of control modes before $\text{BS}_3$ extracted using Eq.~\eqref{eq:compBS3}.

We implemented three single-mode gates, namely the rotation gate, $\hat{R}(\theta)=e^{-i\theta(\x^2+\p^2)/2}$, the shearing gate $\hat{P}(\sigma)=e^{i\sigma\x^2/2}$, and the squeezing gate $\hat{S}(e^r)=e^{ir(\x\p+\p\x)/2}$, each with symplectic matrices
\begin{equation*}
	\textbf{R}_\theta=\begin{pmatrix}
		\cos\theta & \sin\theta \\ -\sin\theta & \cos\theta
	\end{pmatrix}\quad,\quad
	\textbf{P}_\sigma=\begin{pmatrix}
		1 & 0 \\ \sigma & 1
	\end{pmatrix}\quad,\quad\textbf{S}_r=
	\begin{pmatrix}
		e^{-r} & 0 \\ 0 & e^r
	\end{pmatrix}
\end{equation*}
transforming a quadrature vector $(\x,\p)^T$. $\lbrace\hat{R}(\theta),\hat{S}(e^r)\rbrace$ constitute a universal single-mode Gaussian gate set together with displacements in phase-space \cite{ukai10}, while for GKP-encoded qubits on square grids in phase-space $\lbrace\hat{R}(\pi/2),\hat{P}(1)\rbrace$, together with $\sqrt{\pi}$ displacements in phase-space, leads to a single-mode Clifford gate set in the encoded qubit subspace \cite{gottesman01}. Note, phase-space displacements are ubiquitous in measurement-based quantum computation, but they are simply absorbed into the measurement results.

From Eq.~\eqref{eq:G}, $\hat{R}(\theta)$ is implemented from mode $B,k$ to $B,k+N$ with basis setting
\begin{equation*}
	\begin{pmatrix}
		\theta_{A,k}\\\theta_{B,k}
	\end{pmatrix}_R = \frac{1}{2}\begin{pmatrix}
		\theta-(-1)^w\pi/2\\\theta+(-1)^w\pi/2
	\end{pmatrix}\;,
\end{equation*}
such that $(\theta_{+,k},\theta_{-,k})=(\theta,(-1)^w\pi/2)$. Note that $(-1)^w$ in $\theta_{-,k}$ compensates for $(-1)^w$ in Eq.~\eqref{eq:G}. From Eq.~\eqref{eq:G}, $\hat{P}(\sigma)$ cannot be implemented in a single computation step. However, $\hat{F}^j\hat{P}(\sigma)=\hat{R}(j\pi/2)\hat{P}(\sigma)$ can be implemented in a single computation step, in which $\hat{F}^j$ keeps a GKP-encoded qubit within the qubit subspace for $\sigma=1$, and may be compensated for in a following computation step if necessary. Here, with $j=(-1)^w$, $\hat{F}^j=\hat{F}$ for even wires and $\hat{F}^j=\hat{F}^{-1}=\hat{F}^\dagger$ for uneven wires. $\hat{F}^j\hat{P}(\sigma)$ is implemented from mode $B,k$ to $B,k+N$ with the basis setting
\begin{equation*}
	\begin{pmatrix}
		\theta_{A,k}\\\theta_{B,k}
	\end{pmatrix}_P = \begin{pmatrix}
		0 \\ \pi/2-\arctan(\sigma/2)
	\end{pmatrix}\;,
\end{equation*}
such that $\theta_{+,k}=\theta_{-,k}=\pi/2-\arctan(\sigma/2)$. Finally, $\hat{S}(e^r)$ is implemented from mode $B,k$ to $B,k+N$ with basis setting
\begin{equation*}
	\begin{pmatrix}
		\theta_{A,k}\\\theta_{B,k}
	\end{pmatrix}_S = (-1)^w\arctan e^r\begin{pmatrix}
		-1 \\ 1
	\end{pmatrix}\;,
\end{equation*}
such that $(\theta_{+,k},\theta_{-,k})=(0,(-1)^w2\arctan e^r)$. For each of the implemented gates, the displacement by-product is compensated for in the measurement outcomes using Eq.~\eqref{eq:by_product_D}, while with gate noise factors of $4$ in each quadrature, we expect gate noise variance of $4V_0e^{-2r}$ (with $r$ here being the initial squeezing parameter in the cluster state generation) added to each quadrature, corresponding to $\SI{6}{dB}$ relative to the initial momentum squeezed quadrature variance.

\subsection{Two-mode gate}\label{sec2:two}
\begin{figure}
	\centering
	\includegraphics[width=0.78\textwidth]{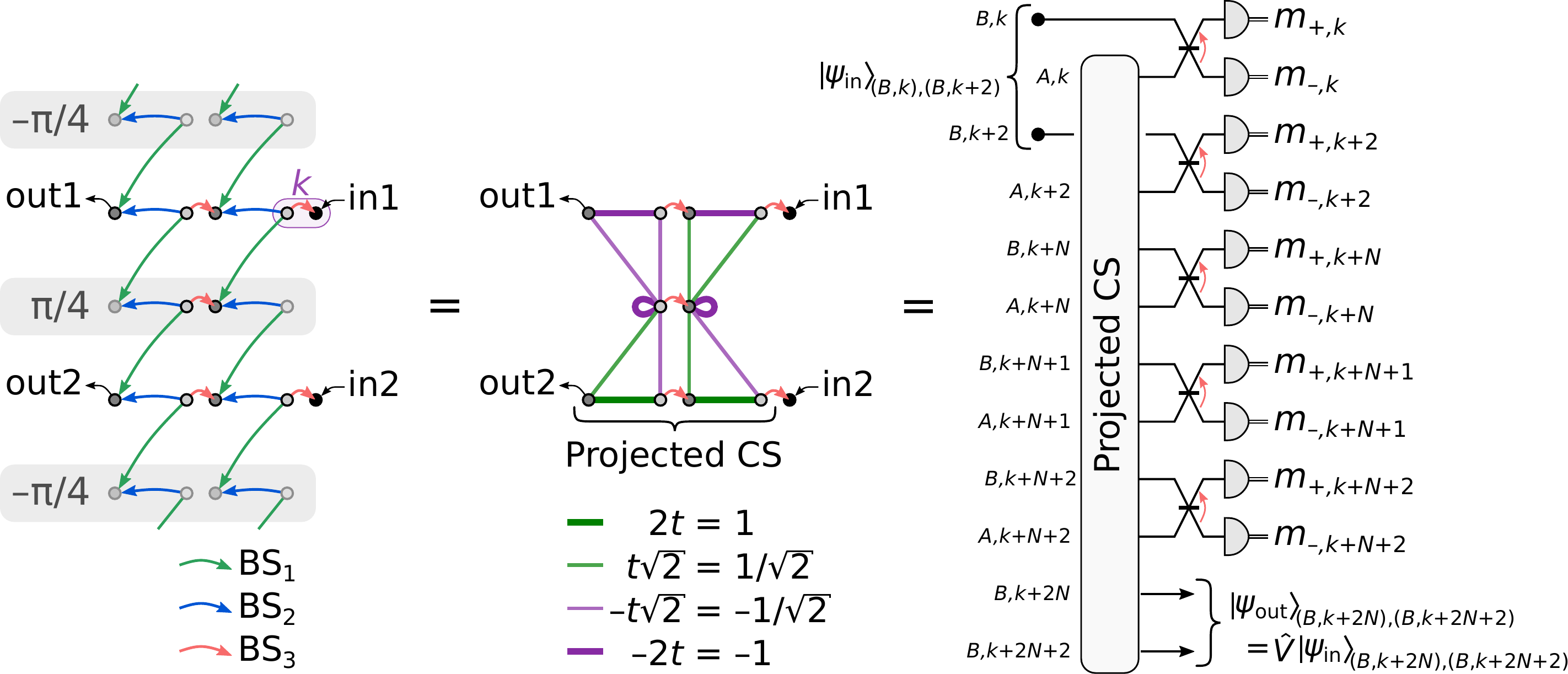}
	\caption{Scheme for implementing two-mode operations. Due to the many modes involved, the circuit for the cluster state generation and projection is shown graphically to the left with each beam splitter represented by an arrow. Here, control modes in the grey shaded areas are phase rotated by $\pm\pi/4$, followed by measuring $\x$ of the faded modes. The resulting projected cluster-type state is shown in the center, consisting of two wires (here an even and odd wire on top and below) and a temporal control mode to couple the wires. By joint measurement of the input using the measurement device with $\text{BS}_3$, as shown to the right, a two-mode operation, $\hat{V}$, with a desired two-mode gate is implemented depending on the basis setting.}
	\label{fig:two_mdoe}
\end{figure}

If, instead of measuring the spatial modes $A$ and $B$ in the same basis for every temporal control mode, we measure them in different bases (such that Eq.~\eqref{eq:compBS3} does not apply), two neighboring wires will be coupled \cite{larsen20}. This can then be used to implement two-mode gates as illustrated in Fig.~\ref{fig:two_mdoe}.

In Fig.~\ref{fig:two_mdoe}, the projected cluster state corresponds to the cluster-type state discussed for wire projection in Fig.~\ref{fig:projection}. However, instead of measuring all control modes in the $\pm\pi/4$-basis, one temporal control mode is only phase-rotated by $(-1)^{(k-1)/2}\pi/4$ but is left unmeasured. The result is a 10-mode projected cluster state, which is then connected to two input modes, $B,k$ and $B,k+2$, via the joint measurement of the measurement device. Measuring all modes, except the output modes $B,k+2N$ and $B,k+2N+2$, the states of the input modes are teleported to the output modes with a two-mode Gaussian operation, $\hat{V}$, applied depending on the basis settings,
\begin{equation*}
	\ket{\psi_\text{out}}_{(B,k+2N),(B,k+2N+2)}=\hat{V}\ket{\psi_\text{in}}_{(B,k),(B,k+2)}\;.
\end{equation*}
In general, $\hat{V}$ transforms the quadratures of the input to the output modes as
\begin{equation}\label{eq:V}
	\begin{pmatrix}
		\x'_{B,k+2N}\\\x'_{B,k+2N+2}\\\p'_{B,k+2N}\\\p'_{B,k+2N+2}\\
	\end{pmatrix}=\textbf{G}_2\begin{pmatrix}
		\x_{B,k}\\\x_{B,k+2}\\\p_{B,k}\\\p_{B,k+2}\\
	\end{pmatrix}+\textbf{N}\boldsymbol{\hat{p}_i}+\textbf{D}\boldsymbol{m}\;,
\end{equation}
where, again, prime marks the output mode quadratures. Here $\textbf{G}_2$ is the symplectic matrix corresponding to the implemented desired two-mode Gaussian gate, $\hat{G}_2$. $\textbf{N}\boldsymbol{\hat{p}_i}$ is the gate noise term with $\textbf{N}$ being a gate noise matrix and $\boldsymbol{\hat{p}_i}$ being a vector of initial momentum squeezed quadratures in the cluster state generation. Finally, $\textbf{D}\boldsymbol{m}$ is the displacement by-product with $\textbf{D}$ being a displacement matrix and $\boldsymbol{m}$ being a vector of measurement outcomes. Here, to shorten the discussion, instead of writing the quadrature transformation initially in terms of the projected cluster state quadratures and following measurement outcomes as for the single-mode operations in Eq.~\eqref{eq:U} (i.e. in terms of $\p_{A(B),k}$ and $\tilde{m}_{\pm,k}$), in the following we will write the quadrature transformation directly from the initial momentum squeezed quadratures and measurement outcomes including the cluster state projection, similar to Eq.~\eqref{eq:N} and \eqref{eq:by_product_D}.

The two-mode controlled-Z gate, $\hat{C}_Z(g)=e^{ig\x\otimes\x}$ with $g$ being a coupling constant, constitutes, together with the single-mode gate set $\lbrace\hat{R}(\theta),\hat{S}(e^r)\rbrace$ in section \ref{sec2:single}, a universal multi-mode Gaussian gate set. Furthermore, for GKP-encoded qubits on square grids in phase-space, $\hat{C}_Z(1)$ leads, together with $\lbrace\hat{F}=\hat{R}(\pi/2),\hat{P}(1)\rbrace$, to a multi-mode Clifford gate set in the encoded qubit subspace \cite{gottesman01}. However, on the cluster state architecture considered here, $\hat{C}_Z(g)$ cannot be implemented directly. Instead, we implement $(\F\otimes\F^j)\hat{C}_Z(g)$, with short notation $\F\F^j\CZ(g)$, where $j=(-1)^w$ with $w$ being the wire number of the input mode $B,k$ (again $\F^{+1}=\F$ and $\F^{-1}=\F^\dagger$). Here, $\hat{F}$ and $\F^j$ on the two output modes keep a GKP-encoded qubit within the encoded qubit subspace for $g=1$, and may be compensated in a subsequent computation step if necessary. The basis setting to implement $\F\F^j\CZ(g)$ from modes $B,k$ and $B,k+2$ to modes $B,k+2N$ and $B,k+2N+2$ is
\begin{equation}\label{eq:CZbasis}
	\begin{pmatrix}
		\theta_{A,k}\\\theta_{B,k}\\\theta_{A,k+2}\\\theta_{B,k+2}\\
		\theta_{A,k+N}\\\theta_{B,k+N}\\\theta_{A,k+N+1}\\\theta_{B,k+N+1}\\\theta_{A,k+N+2}\\\theta_{B,k+N+2}\\
	\end{pmatrix}=
	\begin{pmatrix}
		\pi/4\\-\pi/4\\(-1)^w\pi/4\\-(-1)^w\pi/4\\(-1)^w[\pi/2-\arctan(g/2)]\\0\\(-1)^w\pi/4\\(-1)^w[\pi/4+2\arctan(g/2)]\\(-1)^w[\pi/2-\arctan(g/2)]\\0
	\end{pmatrix}\;.
\end{equation}
This basis setting includes the settings for the temporal control mode $k+N+1$ coupling the two wires $w$ and $w+1$ for the desired two-mode gates, while all other control modes surrounding the two wires are measured in the regular $\pm\pi/4$ control mode basis as described at the beginning of section \ref{sec2}.

\begin{sidewaysfigure}
	\centering
	\includegraphics[width=0.91\textwidth]{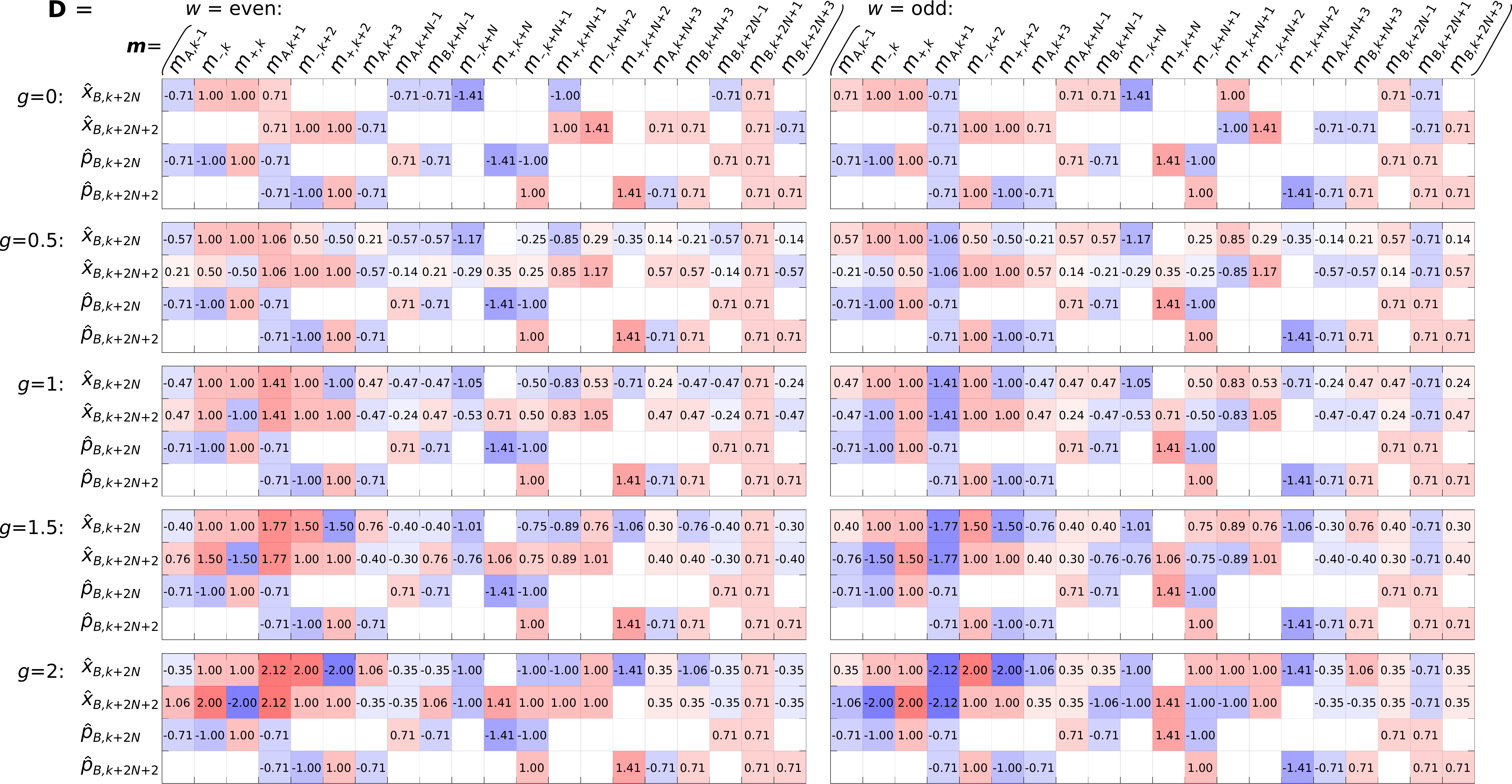}
	\caption{Displacement matrix \textbf{D} in Eq.~\eqref{eq:V} for implementing the $(\F\otimes\F^j)\hat{C}_Z(g)$-gate. Here the left and right column of matrices show \textbf{D} for input mode $B,k$ being on an even and odd wire respectively, while each row represents a different value of coupling constant $g$. Indices without numbers are zero. On top of each column, the vector $\boldsymbol{m}$ of measurement outcomes are shown.}
	\label{fig:D}
\end{sidewaysfigure}

The displacement matrix, \textbf{D}, and corresponding vector of measurement outcomes, $\boldsymbol{m}$, is shown in Fig.~\ref{fig:D} for the different coupling constants $g$ implemented. For the measurement outcomes in $\boldsymbol{m}$ corresponding to the modes in Eq.~\eqref{eq:CZbasis}, the direct homodyne measurement outcomes, $m_{\pm,k}$, are used, while for control modes used to form the projected cluster state in Fig.~\ref{fig:two_mdoe}, the measurement outcomes before $\text{BS}_3$, $m_{A(B),k}$, are used, extracted using Eq.~\eqref{eq:compBS3}. In the characterization of the gate, the displacement is compensated for in the measurement outcomes of the output modes $B,k+2N$ and $B,k+2N+2$.

\begin{sidewaysfigure}
	\centering
	\includegraphics[width=0.89\textwidth]{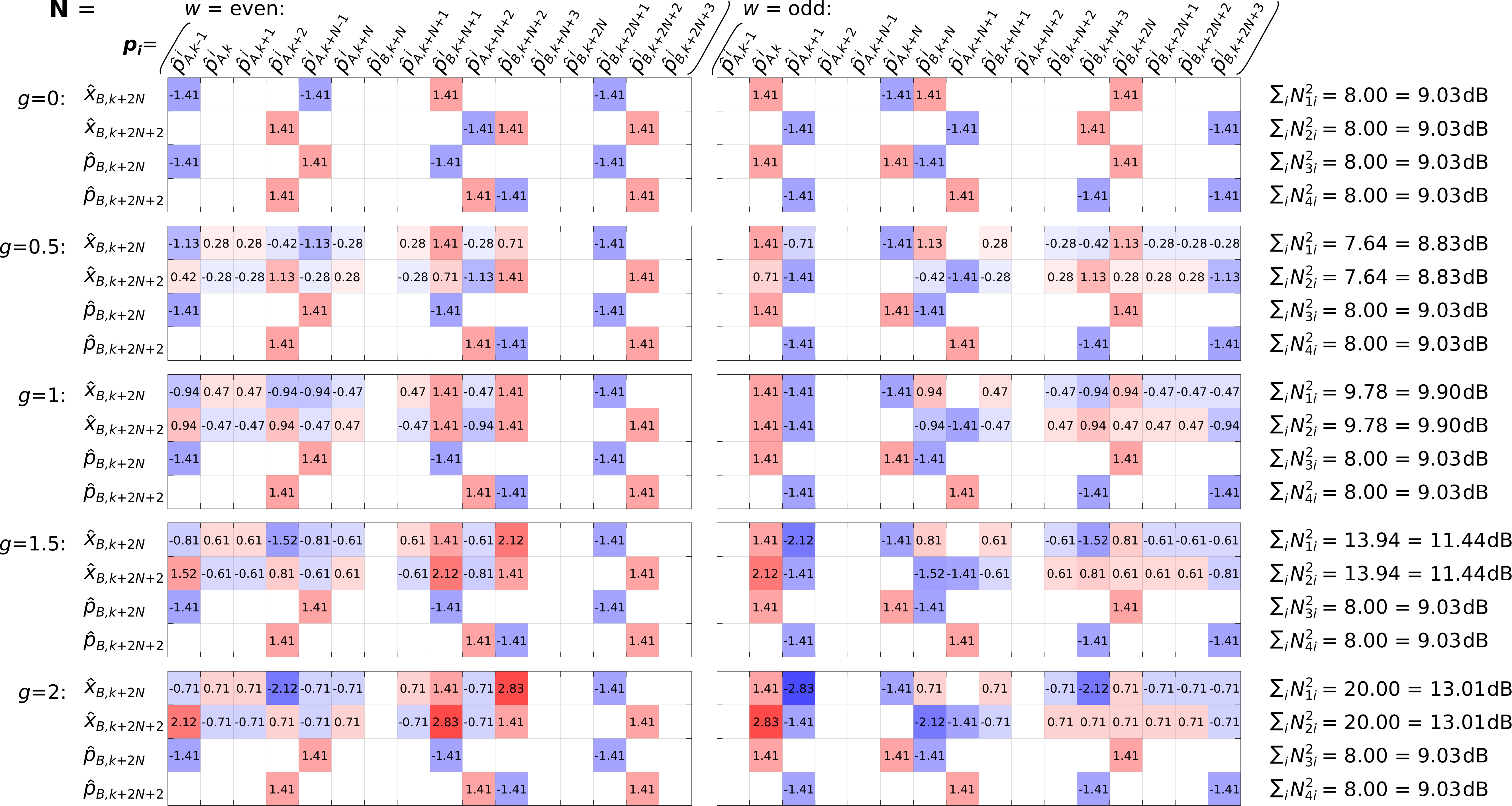}
	\caption{Gate noise matrix \textbf{N} in Eq.~\eqref{eq:V} for implementing the $(\F\otimes\F^j)\hat{C}_Z(g)$-gate. Here the left and right column of matrices show \textbf{N} for input mode $B,k$ being on an even and odd wire, respectively, while each row represents a different value of coupling constant $g$. Indices without numbers are zero. On the very right, the total gate noise factor is shown for each output quadrature, and is equal for the left and right column of gate noise matrices. Above the matrices, the vector $\boldsymbol{p_i}$ of initial momentum squeezed quadratures is shown.}
	\label{fig:N}
\end{sidewaysfigure}

Finally, the gate noise matrix, \textbf{N}, and corresponding vector of initial momentum squeezed quadratures, $\boldsymbol{\hat{p}_i}$, is shown in Fig.~\ref{fig:N}. Assuming all initial momentum squeezed quadratures, $\p_{A(B),k}^i$, are equally squeezed by $e^{-r}$ such that the variances of the momentum quadrature distributions are $e^{-2r}V_0$, gate noise of variance $e^{-2r}V_0\sum_i N_{qi}^2$ is added in each output quadrature. Here, again, $\sum_i N_{qi}^2$ are gate noise factors scaling the initial momentum squeezing variance into gate noise variance with $q=1$, $2$, $3$, and $4$ for the output quadratures $\x_{B,k+2N}$, $\x_{B,k+2N+2}$, $\p_{B,k+2N}$ and $\p_{B,k+2N+2}$, respectively. The resulting gate noise factors are shown in Fig.~\ref{fig:N}.

\subsection{Circuit}\label{sec2:circuit}
Besides implementing individual gates of a multi-mode universal Gaussian gate set, the cluster state computation architecture in this work allows quantum gates to be combined universally, with single-mode gates on each wire by teleportation as described in section \ref{sec2:single}, and two-mode gates between neighboring wires coupled by control modes as described in section \ref{sec2:two}. To demonstrate this, we implement a quantum circuit consisting of 10 single-mode gates and 2 two-mode gates distributed on 3 modes as shown in the main text Fig.~4b. For GKP-encoded qubits encoded on square grids in phase-space, the implemented circuit corresponds to the encoding scheme of a logical qubit in the 3-qubit bit-flip error correction code as shown in the main text Fig.~4a.

\begin{sidewaysfigure}
	\centering
	\includegraphics[width=1\textwidth]{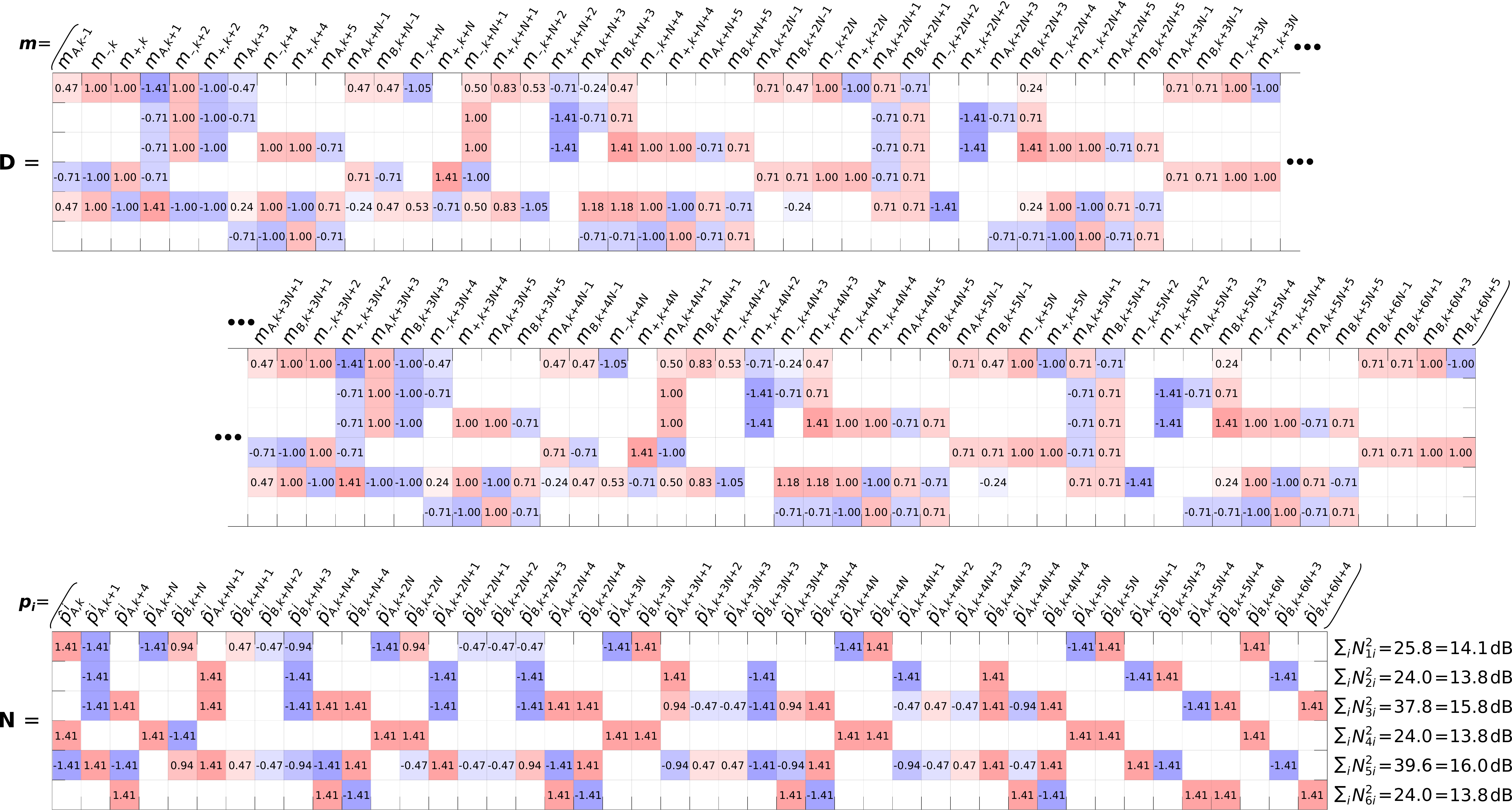}
	\caption{Upper part shows the displacement matrix, $\textbf{D}$, and vector of measurement outcomes, $\boldsymbol{m}$, of the accumulated displacement by-products on the circuit's output quadratures. The matrix rows correspond to the output mode quadratures $\x_{B,k+6N}$, $\x_{B,k+6N+2}$, $\x_{B,k+6N+4}$, $\p_{B,k+6N}$, $\p_{B,k+6N+2}$, and $\p_{B,k+6N+4}$. Similarly, the lower part shows the gate/circuit noise matrix, $\textbf{N}$, and vector of initial momentum squeezed quadratures, $\boldsymbol{p_i}$, of accumulated gate noise on the circuit's output quadratures.}
	\label{fig:circuitDN}
\end{sidewaysfigure}

The circuit is implemented with the three input states on mode $(B,k)$, $(B,k+2)$ and $(B,k+4)$, and output states on mode $(B,k+6N)$, $(B,k+6N+2)$ and $(B,k+6N+4)$. Since in this work the displacement by-product is compensated for in the measurement outcomes, the displacement of each gate will accumulate throughout the computation. By keeping track of each measurement outcome, and how they propagate through the subsequent gates towards the output modes, the displacements are compensated for in the measurements of the circuit's three output states. Here, using the expressions for displacements in single- and two-mode gates in Eq.~\eqref{eq:D} and in Fig.~\ref{fig:D} for $g=1$, an accumulated displacement term, $\textbf{D}\boldsymbol{m}$, on the circuit output modes is derived, and the resulting displacement matrix, \textbf{D}, and vector of measurement outcomes, $\boldsymbol{m}$, are shown in Fig.~\ref{fig:circuitDN}.

Each gate of the implemented circuit adds gate noise to the quadratures of the modes in computation proportional to the initial momentum squeezing in the cluster state generation. To avoid gate noise accumulation, the noise must be removed after each gate implementation. This can be done by using GKP-encoded qubits for computation and applying GKP error correction of the computation mode quadratures \cite{gottesman01} as often as possible, preferably after each implemented gate depending on the magnitude of the gate noise. In the work here, however, GKP-encoding and error correction is not implemented and as a result, the gate noise accumulates during the computation and thus adds to the circuit output modes. This noise is measured and compared to the theoretically expected accumulated gate noise which depends on the initial momentum squeezing. Using the expression for single- and two-mode gate noise in Eq.~\eqref{eq:N} and in Fig.~\ref{fig:N} for $g=1$, and keeping track of the gate noise propagation through each gate of the implemented circuit, a combined circuit noise term of accumulated gate noise, $\textbf{N}\boldsymbol{\hat{p}_i}$, is derived with the noise matrix, \textbf{N}, and the vector of initial momentum squeezed quadrature, $\boldsymbol{\hat{p}_i}$, presented in Fig.~\ref{fig:circuitDN}. The resulting noise factors, $\sum_j N_{qj}^2$, of the output quadratures $\x_{B,k+6N}$, $\x_{B,k+6N+2}$, $\x_{B,k+6N+4}$, $\p_{B,k+6N}$, $\p_{B,k+6N+2}$, and $\p_{B,k+6N+4}$ for $q=1,2,3,4,5,$ and $6$ are also shown. Note that the large circuit noise is due to the lack of error correction. In fault-tolerant quantum computation the purpose of error correction is to prevent gate noise to accumulate into such large values.

\section{Gate tomography}\label{sec3}
In section \ref{sec2} we described how quantum gates and circuits are implemented by projective measurements on the generated cluster state presented in section \ref{sec:setup}. Here, we describe the method of gate and circuit characterization which is based on gate tomography. The procedure is based on probing the gates using entangled states as input and subsequently determine the quadrature correlations that remain after the gate operation. This method was also used in Ref. \cite{asavanant20} but here we generalize it to multi-mode gates and circuits.

\begin{figure}
	\centering
	\includegraphics[width=0.4\textwidth]{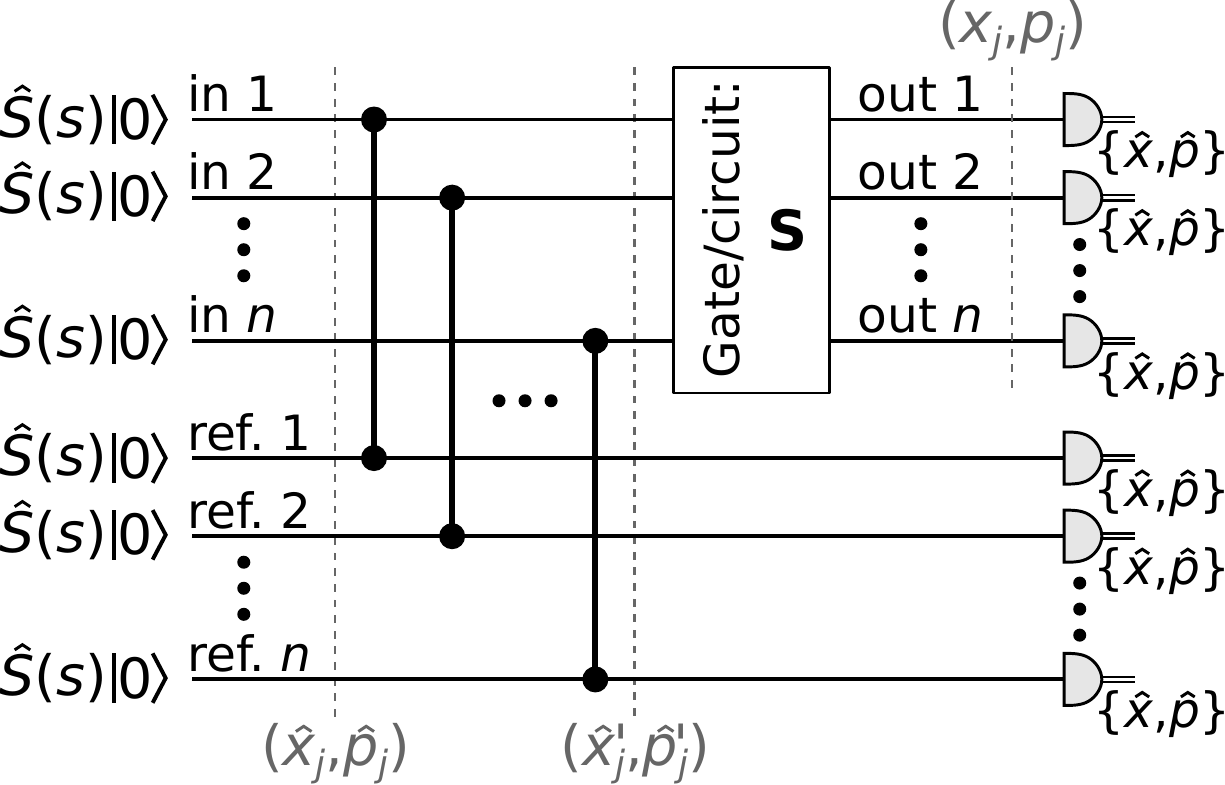}
	\caption{Corresponding circuit of the gate tomography. Here the controlled-Z gates applied to the input and reference modes correspond to the projected two-mode wire cluster states in Fig.~\ref{fig:projection}, where $s$ is a combined squeezed coefficient including additional noise from the wire projection. By measuring the output and reference modes in $\x$- and $\p$-quadrature, the implemented gate/circuit symplectic matrix $\textbf{S}$ can be extracted using Eq.~\eqref{eq:S}, while the gate noise can be estimated from the quadrature variances of the input and output modes. The primed and non-primed quadratures used in the text are marked with a dashed line.}
	\label{fig:tomography}
\end{figure}

The implemented $n$-mode Gaussian gate/circuit, or operation, can be represented by a $2n\times2n$ symplectic matrix $\textbf{S}$ (referred to in section \ref{sec2:single} and \ref{sec2:two} as $\textbf{G}$ and $\textbf{G}_2$) plus gate noise. Here, we consider the case where the displacement by-products have been compensated for, either by feed-forward or, as in this work, in the measurement results. Thus, the goal is to estimate \textbf{S} and the variance of the gate noise in each quadrature. To do so, each of the $n$ input modes to the implemented operation is entangled to one of $n$ reference modes, and from measuring correlations of the reference modes and the operation's output modes we can extract $\textbf{S}$, while the gate noise variance can be estimated from measured variances of the input and output modes together with the estimated $\textbf{S}$. The corresponding circuit is presented in Fig.~\ref{fig:tomography}.

The entanglement probes in Fig.~\ref{fig:tomography} are prepared by the wire projections shown in Fig.~\ref{fig:projection}, and the resulting two-mode entangled states are formed as two-mode wire cluster states. One mode of each of the $n$ two-mode cluster state wires serves as an input mode to the implemented $n$-mode operation, while the other mode serves as the corresponding reference. The projected wires are two-mode squeezed states obtained by interfering two squeezed states, as shown in Fig.~\ref{fig:projection}d, with the corresponding quadrature transformations
\begin{equation*}
	\includegraphics[width=0.61\textwidth]{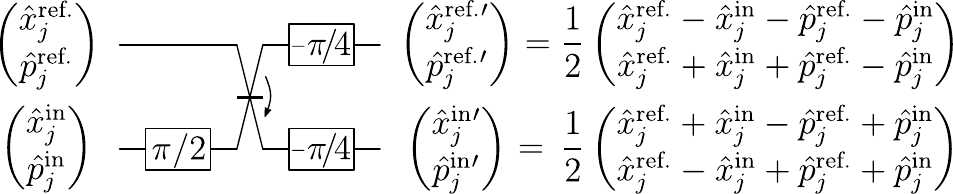}\;,
\end{equation*}
for a $w=\text{odd}$ wire, with a similar transformation for an even wire. Here, the displacement, $\mathcal{D}$, is compensated for, while the $\sqrt{2}$ momentum anti-squeezing contribution and the effect of $\mathcal{N}$ (after moving $\mathcal{N}$ to the left of the beam-splitter and phase shifts in Fig.~\ref{fig:projection}d where it adds uncorrelated noise in connected wire modes as discussed in section \ref{sec2}) are combined in an effective squeezing of $s>e^{-r}$ of the initial input and reference mode momentum quadratures such that 
\begin{equation*}\begin{aligned}
	\braket{(\x_j^\text{ref.})^2}=\braket{(\x_j^\text{in})^2}&=aV_0/s^2\;,\\
	\braket{(\p_j^\text{ref.})^2}=\braket{(\p_j^\text{in})^2}&=s^2V_0\;,
\end{aligned}\end{equation*}
$V_0=1/2$ is the variance of vacuum for $\hbar=1$, and $a=1$ for pure states while excess anti-squeezing is described by $a>1$. Note that here we assume the reference and input modes to be equally squeezed, while the exact values of $s$ and $a$ are unimportant for the purpose here. The resulting quadrature correlations between the input and reference modes are
\begin{equation}\begin{aligned}\label{eq:wireCorr}
\braket{\x_i^\text{in}{}'\x_j^\text{ref.}{}'}=\braket{\p_i^\text{in}{}'\p_j^\text{ref.}{}'}&=\braket{\p_i^\text{in}{}'\x_j^\text{ref.}{}'}=\braket{\x_i^\text{in}{}'\p_j^\text{ref.}{}'}=0\quad,\quad i\neq j\\[5pt]
	\braket{\x_j^\text{in}{}'\x_j^\text{ref.}{}'}=\braket{\p_j^\text{in}{}'\p_j^\text{ref.}{}'}&=0\\
	\braket{\p_j^\text{in}{}'\x_j^\text{ref.}{}'}=\braket{\x_j^\text{in}{}'\p_j^\text{ref.}{}'}&=\frac{1}{2}\left(\frac{a}{s^2}-s^2\right)V_0\equiv \varepsilon_j>0
\end{aligned}\end{equation}
for odd wires with positive edge weight, while similar for even wires with opposite sign of $\varepsilon_j$. Here the first line shows quadrature correlations between input and reference modes of different two-mode wire states which are naturally zero.

After the gate/circuit operation the input mode quadratures are transformed by the symplectic matrix $\textbf{S}$ with entries $s_{i,j}$ while gate noise is added to the output quadratures as
\begin{equation}\label{eq:input_transformation}
	\begin{pmatrix}
		\x_1^\text{out}\\\vdots\\\x_n^\text{out}\\\p_1^\text{out}\\\vdots\\\p_n^\text{out}
	\end{pmatrix}=
	\begin{pmatrix}
		s_{1,1} & \cdots & s_{1,n} & s_{1,n+1} & \cdots & s_{1,2n}\\
		\vdots & \ddots & \vdots & \vdots & \ddots & \vdots \\
		s_{n,1} & \cdots & s_{n,n} & s_{n,n+1} & \cdots & s_{n,2n}\\
		s_{n+1,1} & \cdots & s_{n+1,n} & s_{n+1,n+1} & \cdots & s_{n+1,2n}\\
		\vdots & \ddots & \vdots & \vdots & \ddots & \vdots \\
		s_{2n,1} & \cdots & s_{2n,n} & s_{2n,n+1} & \cdots & s_{2n,2n}
	\end{pmatrix}
	\begin{pmatrix}
		\x_1^\text{in}{}'\\\vdots\\\x_n^\text{in}{}'\\\p_1^\text{in}{}'\\\vdots\\\p_n^\text{in}{}'
	\end{pmatrix}+\textbf{N}\boldsymbol{\hat{p}_i}\;.
\end{equation}
Here the gate noise vector of the initial momentum squeezed quadrature vector, $\boldsymbol{\p_i}$, transformed by the gate noise matrix, $\textbf{N}$, is described in section \ref{sec2}. Using Eq.~\eqref{eq:wireCorr} the quadrature correlations between the input and reference modes can then be used to extract $s_{i,j}$ since
\begin{equation*}\begin{aligned}
	\braket{\x_i^\text{out}\p_j^\text{ref.}{}'}&=\braket{\left(s_{i,1}\x_1^\text{in}{}'+\cdots + s_{i,n}\x_n^\text{in}{}'+s_{i,n+1}\p_1^\text{in}{}'+\cdots + s_{i,2n}\p_n^\text{in}{}'\right)\p_j^\text{ref.}{}'}\\
	&=s_{i,j}\braket{\x_j^\text{in}{}'\p_j^\text{ref.}{}'}+s_{i,n+j}\braket{\p_j^\text{in}{}'\p_j^\text{ref.}{}'}\\
	&=s_{i,j}\varepsilon_j
\end{aligned}\end{equation*}
\begin{equation*}\begin{aligned}
	\braket{\x_i^\text{out}\x_j^\text{ref.}{}'}&=\braket{\left(s_{i,1}\x_1^\text{in}{}'+\cdots + s_{i,n}\x_n^\text{in}{}'+s_{i,n+1}\p_1^\text{in}{}'+\cdots + s_{i,2n}\p_n^\text{in}{}'\right)\x_j^\text{ref.}{}'}\\
	&=s_{i,j}\braket{\x_j^\text{in}{}'\x_j^\text{ref.}{}'}+s_{i,n+j}\braket{\p_j^\text{in}{}'\x_j^\text{ref.}{}'}\\
	&=s_{i,n+j}\varepsilon_j
\end{aligned}\end{equation*}
\begin{equation*}\begin{aligned}
	\braket{\p_i^\text{out}\p_j^\text{ref.}{}'}&=\braket{\left(s_{n+i,1}\x_1^\text{in}{}'+\cdots + s_{n+i,n}\x_n^\text{in}{}'+s_{n+i,n+1}\p_1^\text{in}{}'+\cdots + s_{n+i,2n}\p_n^\text{in}{}'\right)\p_j^\text{ref.}{}'}\\
	&=s_{n+i,j}\braket{\x_j^\text{in}{}'\p_j^\text{ref.}{}'}+s_{n+i,n+j}\braket{\p_j^\text{in}{}'\p_j^\text{ref.}{}'}\\
	&=s_{n+i,j}\varepsilon_j
\end{aligned}\end{equation*}
\begin{equation*}\begin{aligned}
	\braket{\p_i^\text{out}\x_j^\text{ref.}{}'}&=\braket{\left(s_{n+i,1}\x_1^\text{in}{}'+\cdots + s_{n+i,n}\x_n^\text{in}{}'+s_{n+i,n+1}\p_1^\text{in}{}'+\cdots + s_{n+i,2n}\p_n^\text{in}{}'\right)\x_j^\text{ref.}{}'}\\
	&=s_{n+i,j}\braket{\x_j^\text{in}{}'\x_j^\text{ref.}{}'}+s_{n+i,n+j}\braket{\p_j^\text{in}{}'\x_j^\text{ref.}{}'}\\
	&=s_{n+i,n+j}\varepsilon_j
\end{aligned}\end{equation*}
for $i,j\in\{1,\cdots,n\}$. Thus, finally, the symplectic matrix of the implemented operation can be estimated from quadrature correlations of output and reference modes as
\begin{equation}\label{eq:S}
	\textbf{S}=
	\left(
		\begin{array}{c c c|c c c}
			 & \vdots &  &  & \vdots & \\
			\cdots & \braket{\x_i^\text{out}\p_j^\text{ref.}{}'}/\varepsilon_j & \cdots & \cdots & \braket{\x_i^\text{out}\x_j^\text{ref.}{}'}/\varepsilon_j & \cdots\\
			 & \vdots &  &  & \vdots & \\
			\hline
			 & \vdots &  &  & \vdots & \\
			\cdots & \braket{\p_i^\text{out}\p_j^\text{ref.}{}'}/\varepsilon_j & \cdots & \cdots & \braket{\p_i^\text{out}\x_j^\text{ref.}{}'}/\varepsilon_j & \cdots\\
			 & \vdots &  &  & \vdots & \\
		\end{array}
	\right)
\end{equation}
Here, $\varepsilon_j$ is estimated from the measurements of the quadrature correlations between input and reference modes on even and odd wire states individually side by side with the implemented operations. As such, within even and odd wires, we assume $\varepsilon_j$ to be identical for different $j$. This is a valid assumption, as the gate tomography of each implemented gate (including estimating $\varepsilon_j$ for even and odd wires) in this work is carried out within $19N=228$ temporal modes, i.e. within a time period of $228\times\SI{247}{ns}=\SI{56.3}{\micro s}$, while the setup stability allows for a stable setup without changes for much longer time periods.

Using the estimated symplectic matrix in Eq.~\eqref{eq:S}, the variance of the gate noise in Eq.~\eqref{eq:input_transformation} can be estimated as the difference of the quadrature variance on the output modes and the expected quadrature variance from the input modes as
\begin{equation}\label{eq:GN}
	\text{Var}\{\textbf{N}\boldsymbol{\hat{p}_i}\}=	
	\text{Var}\left\lbrace
	\begin{pmatrix}
		\x_1^\text{out}\\\vdots\\\x_n^\text{out}\\\p_1^\text{out}\\\vdots\\\p_n^\text{out}
	\end{pmatrix}
	\right\rbrace-\text{Var}\left\lbrace\textbf{S}
	\begin{pmatrix}
		\x_1^\text{in}{}'\\\vdots\\\x_n^\text{in}{}'\\\p_1^\text{in}{}'\\\vdots\\\p_n^\text{in}{}'
	\end{pmatrix}
	\right\rbrace\;.
\end{equation}
Here the reference mode is traced out. Similar to $\varepsilon_j$, $\text{Var}\{\x_j^\text{in}{}'\}$ and $\text{Var}\{\p_j^\text{in}{}'\}$ are estimated by measurements on wire cluster state individually for even and odd wires, both before and after the gate tomography, but within $19N=228$ temporal modes.

\subsection{Single-mode gates}\label{sec3:single}
\begin{figure}
	\centering
	\includegraphics[width=0.95\textwidth]{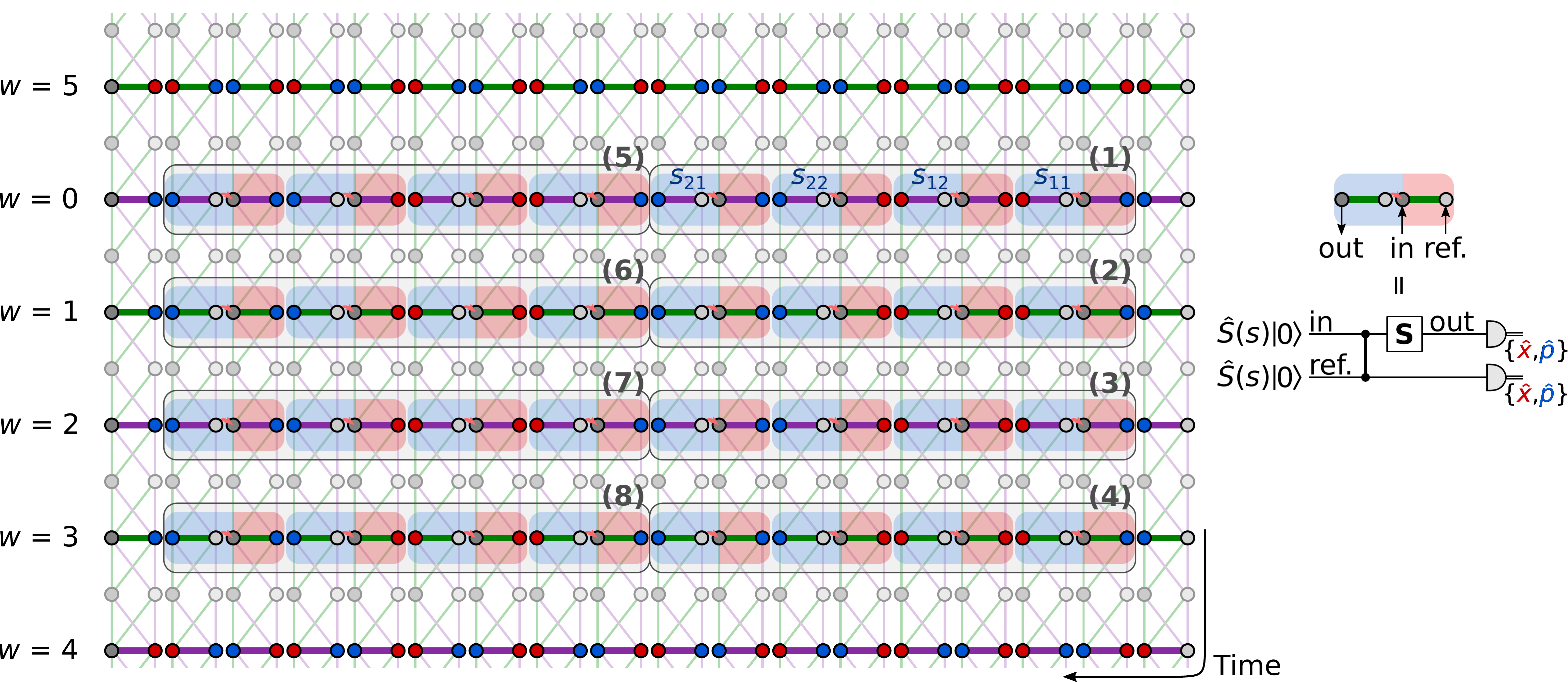}
	\caption{Basis setting layout with $19N=228$ temporal modes for implementation and characterization by gate tomography of 8 different single-mode gates on wire $w=0$--$3$, each in the regions marked by (1--8). In each region the gate is implemented 4 times in order to measure $px$-, $xx$-, $xp$- and $pp$-correlations between the reference and output modes to estimate the indices $s_{1,1}$, $s_{1,2}$, $s_{2,2}$ and $s_{2,1}$, respectively, of the implemented gate symplectic matrix $\textbf{S}$ by Eq.~\eqref{eq:S}. Here, modes marked by red and blue are measured in $\x$- and $\p$-basis, respectively, while faded modes are control modes measured in the $\pm\pi/4$ basis to form wires. The faded edges to control modes are only shown to indicate the initial coiled up 1D cluster state before wire projection, and has now physical meaning here. Even though present in all temporal modes, the beam splitter $\text{BS}_3$ is only shown between spatial modes measured in different basis in order to implement the desired gates. The measurement outcomes before $\text{BS}_3$ on all other modes are extracted using Eq.~\eqref{eq:compBS3}. Wire $w=4$ and $5$ are used to estimate the quadrature correlations between input and reference modes, $\varepsilon_j$, together with the quadrature variance of input modes in order to estimate the gate noise variance by Eq.~\eqref{eq:GN}. To the right of the cluster state basis layout, the corresponding single-mode gate tomography circuit of input, output, and reference modes is shown.}
	\label{fig:tomography_single}
\end{figure}
In Fig.~\ref{fig:tomography_single} the basis setting layout on the generated cluster state for single-mode gate implementation and tomography is shown. Since $\textbf{S}$ is a $2\times2$ matrix for single-mode gates, 4 different measurements of quadrature correlations of the output-reference mode are required to estimate $\textbf{S}$. Thus, with a section of $19N=228$ temporal modes, 8 different single-mode gates can be implemented and characterized at once in parallel on 4 wires. The remaining 2 of the 6 wires are used to estimate the correlations between input and reference modes, $\varepsilon_j$, of the two-mode wire states in both odd and even wires, together with the input mode variance when tracing out the reference modes. To build up statistics, measurements with the layout in Fig.~\ref{fig:tomography_single} are repeated $\SI{10000}{}$ times. In the following we will refer to such $\SI{10000}{}$ repeated measurements of the same layout as a \textit{set} of measurements.

Each of the implemented single-mode gates, $\R(\theta)$, $\hat{P}(\sigma)$ and $\Sq(e^{-r})$, are implemented in different sets of measurements with 7 different values of $\theta$, $\sigma$ and $r$. To fill out one set, which can implement and characterize 8 different gates, the value $\theta,\sigma,r=0$ are repeated twice in each set. From wire $w=4$ in each set the input-reference quadrature correlations are estimated to be $\varepsilon_j=-2.14\pm0.03$, $-2.15\pm0.03$ and $-2.14\pm0.03$ for the sets with $\R(\theta)$, $\hat{P}(\sigma)$ and $\Sq(e^{-r})$, respectively, and are used for gate tomography on $w=\text{even}$ wires with negative edge weight. Similarly, from wire $w=5$, $\varepsilon_j=2.11\pm0.04$, $2.12\pm0.03$ and $2.11\pm0.02$ are estimated and used for gate tomography on $w=\text{odd}$ wires with positive edge weight. The uncertainties on $\varepsilon_j$ are estimated as the standard deviation of the 16 measured quadrature correlations in each wire $w=4$ and $5$ for each measurement set. The resulting symplectic matrix for each implemented single-mode gates, extracted using Eq.~\eqref{eq:S}, are shown in the main text Fig.~2a as a function of $\theta$, $\sigma$ and $r$.

Also, using wires $w=4$ and $5$, the quadrature variances of the input modes when tracing out the reference mode (i.e. ignoring the reference mode measurement outcome) are estimated to be $(\text{Var}\{x_j^\text{in}{}'\},\text{Var}\{p_j^\text{in}{}'\})=(2.28\pm0.04,2.67\pm0.05)$, $(2.33\pm0.03,2.64\pm0.03)$ and $(2.31\pm0.02,2.66\pm0.02)$ on wire $w=4$ for measurement sets with $\R(\theta)$, $\hat{P}(\sigma)$ and $\Sq(e^{-r})$ respectively, and similarly $(\text{Var}\{x_j^\text{in}{}'\},\text{Var}\{p_j^\text{in}{}'\})=(2.30\pm0.03,2.78\pm0.04)$, $(2.34\pm0.03,2.75\pm0.03)$ and $(2.31\pm0.03,2.75\pm0.03)$, respectively, on wire $w=5$. The uncertainties in the variances of the $\x$ and $\p$ quadratures in each measurement set are estimated as standard deviation of the 8 measurements of input modes in each quadrature on wires $w=4$ and $5$. The resulting gate noise variances, extracted using Eq.~\eqref{eq:GN} with the variance of the output modes measured in each gate tomography, are shown in the main text Fig.~2c. Also, the gate noise variance compensated by the theoretical gate noise factor of $\SI{6}{dB}$ is shown, in which case it agrees with the initial momentum squeezing of $\SI{4.4}{dB}$ below the vacuum variance as discussed at the end of section \ref{sec2:single} (a more general discussion of gate noise is presented in section \ref{sec4}).

To show how the gate noise behaves as a function of squeezing, in the main text Fig.~2d the gate noise of $\hat{R}(\theta)$, averaged over $\x$ and $\p$ and over the 7 different implemented values of $\theta$, is shown as a function of pump power relative to the OPO thresholds. The measured gate noise is seen to agree well with the gate noise expected for the estimated properties of the experimental setup and is discussed further in section \ref{sec4}.

\subsection{Two-mode gate}\label{sec3:two}
\begin{figure}
	\centering
	\includegraphics[width=0.96\textwidth]{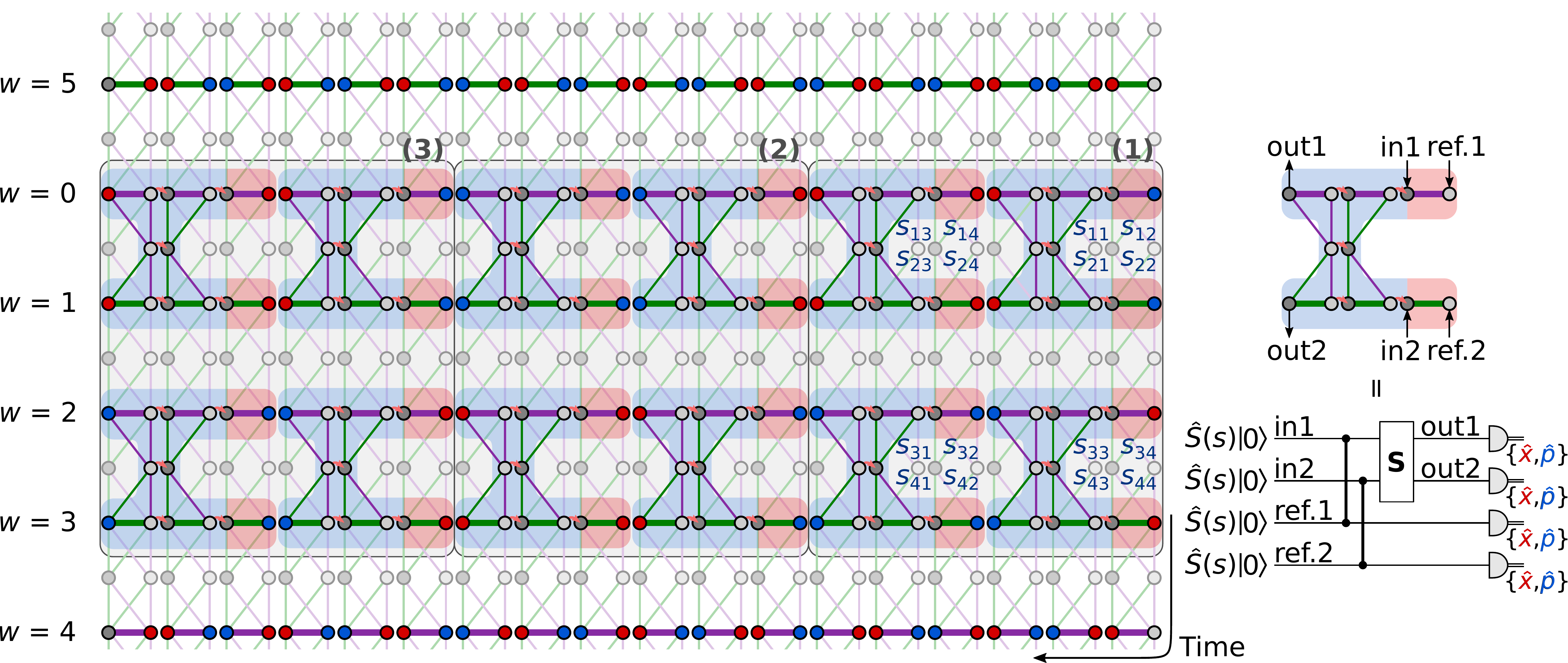}
	\caption{Basis setting layout for implementation and characterization by gate tomography of the two-mode $\CZ(g)$-gate. The notation is the same as in Fig.~\ref{fig:tomography_single} for single-mode gates. The same $\CZ(g)$-gate is implemented 4 times for gate tomography. As a result, 3 different $\CZ(g)$-gates can be implemented and characterized within one measurement set of $19N=228$ temporal modes in the regions marked (1--3). $\CZ(g)$ is implemented and characterized for 5 different values of $g$ in two sets of measurements. To the right, the corresponding circuit of the two-mode gate tomography circuit of input, output, and reference modes is shown.}
	\label{fig:tomography_two}
\end{figure}

When performing gate tomography on the two-mode controlled-Z gate, $\CZ(g)$, the $4\times4=16$ entries $s_{i,j}$ of the corresponding symplectic matrix $\textbf{S}$ can be estimated by executing the $\CZ(g)$ gate four times, since 4 different input-reference mode quadrature correlations can be estimated simultaneously using the two reference and output modes of each implementation. The basis setting layout of $228$ temporal modes is shown in Fig.~\ref{fig:tomography_two}, in which $\CZ(g)$ can be implemented with three different values of $g$. In total, $\CZ(g)$ is implemented with 5 different values of $g$ distributed on two sets of measurements with $g=0$ repeated twice to fill out the sets.

Similar to the gate tomography of single-mode gates in section \ref{sec3:single}, the input-reference quadrature correlations, $\varepsilon_j$, averaged on the two measurement sets, are estimated from wires $w=4$ and $5$ to be $\varepsilon_j=-2.15\pm0.04$ and $2.13\pm0.03$, respectively, with uncertainties estimated as the standard deviation of 16 measured quadrature correlations in each wire. From the gate tomography, the resulting symplectic matrix estimated using Eq.~\eqref{eq:S} is shown in the main text Fig.~3a as a function of $g$.

From wires $w=4$ and $5$ the input mode variances, averaged over the two measurement sets, are estimated to be $(\text{Var}\{x_j^\text{in}{}'\},\text{Var}\{p_j^\text{in}{}'\})=(2.36\pm0.06,2.72\pm0.04)$ and $(2.34\pm0.03,2.65\pm0.04)$ on even and odd wires respectively, with uncertainties estimated as the standard deviation of the 8 measured input modes per $\x$- and $\p$-quadrature of each wire. Using this, and the estimated $\textbf{S}$, the gate noise variance is estimated using Eq.~\eqref{eq:GN}, and shown in the main text Fig.~3c as a function of $g$. Here, the gate noise variance is also compensated for by the theoretical gate noise factor presented in Fig.~\ref{fig:N}, in which case the gate noise agrees well with the initial momentum squeezed variance of $\SI{4.4}{dB}$ below the vacuum variance.

\subsection{Circuit}\label{sec3:circuit}
\begin{figure}
	\centering
	\includegraphics[width=0.85\textwidth]{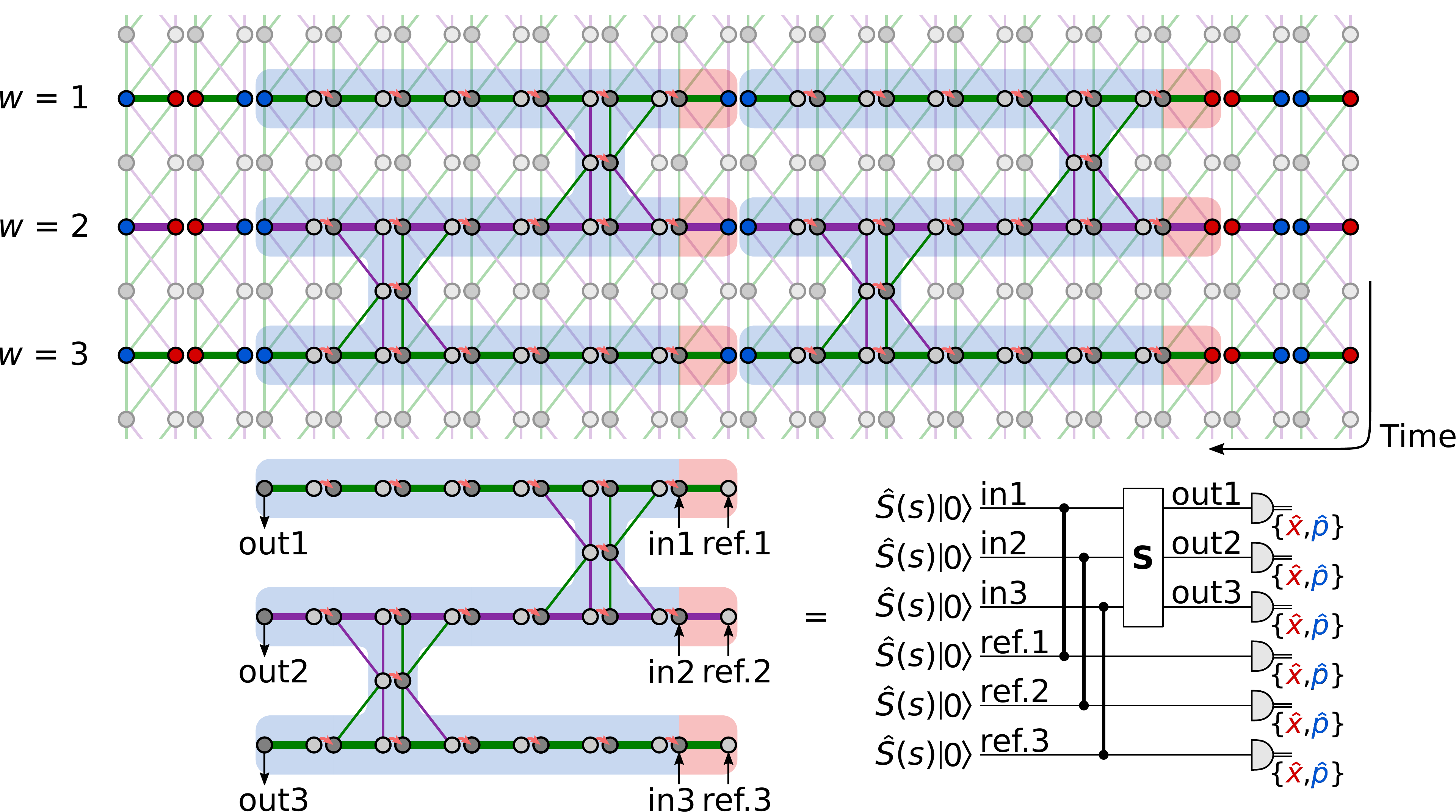}
	\caption{Basis setting layout for implementation and characterization of the circuit described in section \ref{sec2:circuit}, here implemented on wire $w=1$, 2 and 3, while wire $w=0$, 4 and 5 are not shown here. The notation is the same as in Fig.~\ref{fig:tomography_single} for single-mode gates. For full tomography the circuit is implemented 4 times, distributed on two sets of measurements. Here, the first set is shown from which the first 6 columns of the symplectic matrix, \textbf{S}, is estimated. In the second set for estimating the last 6 columns of \textbf{S}, the $\x$- and $\p$-basis settings on all output and reference modes are swapped. Below the basis setting layout, the corresponding circuit of the multi-mode circuit tomography with input, output, and reference modes is shown.}
	\label{fig:tomography_circuit}
\end{figure}

The quantum circuit, described in section \ref{sec2:circuit}, is implemented on wires $w=1$, 2 and 3 as shown in the basis setting layout in Fig.~\ref{fig:tomography_circuit}. Similar to the $\CZ(g)$-gate in section \ref{sec3:two}, all $6\times6=36$ entries of the corresponding symplectic matrix $\textbf{S}$ can be extracted from implementing the circuit 4 times, as $3\times3=9$ indices can be estimated from the three reference and output modes of one implementation. Two measurement sets are required for full gate tomography of the circuit implemented on wires $w=1$, 2 and 3: One as in Fig.~\ref{fig:tomography_circuit} where the first 6 columns of $\textbf{S}$ are estimated, and a similar set where the $\x$- and $\p$-basis settings are swapped on the reference and output modes in order to estimate the last 6 columns of $\textbf{S}$.

The first and last two time steps on all wires, $w=0$--5 (though only wire $w=1$, 2 and 3 are shown in Fig.~\ref{fig:tomography_circuit}), are used to estimate the input-reference mode quadrature correlations, $\varepsilon_j$, and are estimated to be $\varepsilon_j=-2.14\pm0.03$ and $2.12\pm0.06$ on even and odd wires, respectively, averaged over the two measurement sets. Uncertainties are estimated as the standard deviation of 12 measured quadrature correlations each for even and odd wires. The resulting estimated symplectic matrix using Eq.~\eqref{eq:S} is shown in the main text Fig.~4c.

Similar to $\varepsilon_j$, the quadrature variances of input modes are estimated from the first and last two time steps in each wire when tracing out reference modes by ignoring the reference mode measurement outcomes. The input mode quadrature variances are estimated to be $(\text{Var}\{x_j^\text{in}{}'\},\text{Var}\{p_j^\text{in}{}'\})=(2.34\pm0.05,2.72\pm0.03)$ and $(2.34\pm0.04,2.68\pm0.08)$ on even and odd wires respectively. Uncertainties are estimated as the standard deviation from 6 measured input modes per $\x$- and $\p$-quadrature each in even and odd wires. Using this, and the estimated \textbf{S}, the gate noise variance is estimated using Eq.~\eqref{eq:GN}, and shown in the main text Fig.~4d. The gate noise is also shown when compensating for the theoretical gate noise factors presented in Fig.~\ref{fig:circuitDN}, and the resulting gate noise agrees well with the initial momentum squeezed variance of $\SI{4.4}{dB}$ below the vacuum variance.

\section{Gate noise}\label{sec4}
In this section, we discuss the performance with respect to measured gate noise in the cluster state computation scheme of this work. Ideal cluster states have nullifiers of zero variance. However, such ideal cluster states, generated from infinite squeezed states, are non-physical. Instead, as discussed in section \ref{sec2}, we generate approximate cluster states, or cluster-type states, from finitely squeezed states with nullifier variances that vanish in the limit of infinite squeezing but are otherwise non-zero. When implementing gates on the cluster state by projective measurement, the inherent noise in the cluster state from finite squeezing leads to gate noise proportional to the initial squeezing in the cluster state generation as discussed in section \ref{sec2:single} and \ref{sec2:two}---the gate noise variance in each quadrature is the product of the initial squeezing variance and the corresponding gate noise factor.

To prevent gate noise from accumulating on the states in computation, error correction is required. This can be achieved by encoding the computational states as qubits using Gottesman-Kitaev-Preskill (GKP) encoding, where qubits are encoded in the bosonic mode quadratures \cite{gottesman01}. The gate noise, which appears as noise in the quadratures, can then be corrected with GKP quadrature error correction with a success probability depending on the amount of squeezing. In the event of an error in the quadrature error correction, a bit-flip and/or phase-flip error is induced on the encoded qubit, which can then be corrected with a concatenated qubit error correction code. A detailed review of GKP encoding and error correction can be found in \cite{gottesman01,tzitrin20,terhal20}, while in \cite{larsen20,menicucci14} GKP error correction and resulting qubit error probability is reviewed in terms of gate noise. To achieve fault-tolerant quantum computation, the probability of inducing a qubit error on the GKP-encoded qubits should be lower than the error probability threshold allowed by the concatenated qubit error correction code. Since this qubit error probability depends on the gate noise and thereby the initial squeezing, the amount of required squeezing for fault-tolerance depends on the error probability that the concatenated qubit error correction code can tolerate. Different squeezing levels have been estimated in the settings of cluster state computation, ranging from at highest $\SI{20}{dB}$ below the vacuum variance \cite{menicucci14}, to $\SI{17}{dB}$ in \cite{walshe19} where excess anti-squeezing is shown to not affect the squeezing threshold, and further to $\SI{10}{dB}$ in \cite{fukui18} using topological qubit error correction codes (though this requires a 3-dimensional cluster state instead of the 2-dimensional cluster state used in this work).

In the other limit of no squeezing, the generated cluster state is simply a vacuum state. In this case, a desired gate can still be implemented, as the quadrature transformation is mainly realized by the feedforward of the displacement by-product, and as such, this is equivalent to classical teleportation \cite{braunstein98}. In this limit, considering single-mode gates, the gate noise is $4$ units of vacuum noise, $4V_0$, as discussed in section \ref{sec2:single}, i.e. $\SI{6}{dB}$ above the vacuum variance. Though single-mode gates can be implemented with less gate noise in classical teleportation, in the cluster state computation scheme in this work, it is not possible to lower the gate noise below $4V_0=\SI{6}{dB}$ without squeezing in the cluster state generation leading to quantum correlations between the cluster state modes. With $e^{-r}$ momentum quadrature squeezing, the single-mode gate noise becomes $4e^{-2r}V_0$. In \cite{larsen19} the cluster state modes are shown to be inseparable when the cluster state is generated with more than $\SI{3}{dB}$ of squeezing, while in the work here the cluster state is generated with $\SI{4.4}{dB}$ of squeezing, which leads to expected single-mode gate noise of $\SI{-4.4}{dB}+\SI{6}{dB}=\SI{1.6}{dB}=1.4V_0$. This expected gate noise agrees with the measured gate noise shown in the main text Fig.~2c, where the gate noise compensated for by $\SI{6}{dB}$ (hollow points) is compared to the $\SI{4.4}{dB}$ initial squeezing (dashed line). The same analysis is shown for the two-mode $\CZ(g)$-gate and accumulated gate noise of the implemented circuit in the main text Fig.~3c and 4d.

In conclusion, even though the gate noise measured in this work is larger than what can be achieved with different architectures, it is less than what can be achieved with the computation scheme used here without the presence of a cluster state~\cite{larsen20}. In the cluster state computation scheme here, the function of the cluster state can be thought of as to decrease the gate noise depending on the cluster state quality in terms of initial squeezing. In the main text Fig.~2d, the measured single-mode gate noise is shown as a function of pump power. At zero pump power where the cluster state is simply vacuum, we measured $\SI{6}{dB}$ gate noise as expected, while increasing the pump power, generating a cluster state, decreases the gate noise. At some point the gate noise saturates due to optical losses and phase fluctuations, limiting the cluster state quality achievable in the given setup. This is not a fundamental limit  but merely a technical challenge. For fault-tolerant computation, optical losses and phase control need to be optimized to improve the cluster state quality, while the computation scheme will remain the same as demonstrated here. Even though some cluster state architectures, like the quad-rail lattice \cite{menicucci11b,alexander16a}, can offer an improvement in the gate noise by $\SI{3}{dB}$ \cite{larsen20}, to achieve fault-tolerant computation it is not enough to simply measure gate noise less than the vacuum variance, $V_0=\SI{0}{dB}$; optical losses and phase control need to be optimized for any architecture. As such, although the gate noise is a good figure of merit, a gate noise below $V_0$ is not the target---it is gate noise below some fault-tolerant squeezing threshold which is the target. Below we estimate how the gate noise would scale in more optimal settings of the computation scheme in this work.

\subsection{Gate noise in optimal settings}
To estimate the gate noise, it is sufficient to estimate the initial quadrature squeezing in the cluster state generation and multiply with the gate noise factors $\sum_i N_{qi}^2$ with $N_{qi}$ being entries of the gate noise matrix $\textbf{N}$ as discussed in section \ref{sec2}. For single-mode gates, the gate noise factor is $\sum_i N_{1i}^2=\sum_i N_{2i}^2=4$ in both $\x$- and $\p$-quadrature, corresponding to adding $\SI{6}{dB}$ to the initial squeezing, while the gate noise factor for the $\CZ(g)$-gate is given in Fig.~\ref{fig:N}. Thus, we need to first estimate the initial momentum squeezing in the cluster state temporal modes.

The temporal modes are defined by the temporal mode function, $f_k(t)$, in Eq.~\eqref{eq:modeFunction}. For the initial momentum squeezed states in the cluster state generation $\braket{\x}=\braket{\p}=0$, and the quadrature variance of a temporal mode $k$ becomes
\begin{equation}\begin{aligned}\label{eq:modeVar}
	\text{Var}\lbrace \hat{q}_k\rbrace = \braket{\hat{q}^2}&=\left\langle\int f_k(t)\hat{q}(t)\,\text{d}t \int f_k(t')\hat{q}(t')\,\text{d}t'\right\rangle\\
	&=\iint f_k(t)f_k(t')\braket{\hat{q}(t)\hat{q}(t')}\,\text{d}t\,\text{d}t'\quad,\;\hat{q}=\x,\p\;,
\end{aligned}\end{equation}
where $\braket{\x(t)\x(t')}$ and $\braket{\p(t)\p(t')}$ are the quadrature auto-covariance functions. For momentum squeezed states generated as output from an optical parametric oscillator (OPO) pumped below threshold, the quadrature auto-covariance functions are
\begin{equation}\begin{aligned}\label{eq:autoCov}
	\braket{\hat{x}(t)\hat{x}(t')}&=\frac{1}{2}\delta(t-t')+\frac{\eta\gamma\varepsilon}{\gamma-\varepsilon}e^{-(\gamma-\varepsilon)|t-t'|}\;,\\
	\braket{\hat{p}(t)\hat{p}(t')}&=\frac{1}{2}\delta(t-t')-\frac{\eta\gamma\varepsilon}{\gamma+\varepsilon}e^{-(\gamma+\varepsilon)|t-t'|}\;,
\end{aligned}\end{equation}
where $\eta$, $\gamma$, and $\varepsilon$ are the overall efficiency, the OPO decay rate (or bandwidth), and pump rate, respectively \cite{collett84}. In the experimental setup $\gamma=2\pi\times\SI{7.7}{MHz}$ on average for the two OPO squeezing sources \cite{larsen19}.

Inserting Eq.~\eqref{eq:modeFunction} and \eqref{eq:autoCov} into Eq.~\eqref{eq:modeVar}, and using that $\varepsilon=\gamma\sqrt{P/P_\text{thr.}}$ where $P/P_\text{thr.}$ is the pump power relative to the OPO threshold, we estimate the quadrature squeezing and anti-squeezing as a function of pump power for different efficiencies. Adding $\SI{6}{dB}$ to the noise variance, we estimate the resulting single-mode gate noise, which is presented in the main text Fig.~2d for some optimal settings with high efficiencies of $\eta\geq90\%$. Here, the gate noise is presented with two different OPO bandwidths, namely $\gamma=2\pi\times\SI{7.7}{MHz}$ as in the experimental setup, and $\gamma=2\pi\times\SI{100}{MHz}$ for more optimal settings. Predicted gate noise for the current experimental efficiency of $\eta=77.7\%$ (an average of $78.9\%$ and $76.4\%$ of two fitted efficiencies in the supplementary information of \cite{larsen19}), together with an estimated phase fluctuation of $\sigma=4^\circ$, is shown as well and is seen to agree well with the experimentally measured gate noise. The same analysis is possible for the two-mode $\CZ(g)$-gate using the gate noise factors in Fig.~\ref{fig:N}. In general, the gate noise is seen to decrease with increasing efficiencies and pump power. Furthermore, the gate noise is also seen to decrease with increasing squeezing bandwidth. This is explained by a better coverage of the temporal mode frequency response by the wider squeezing bandwidth of $\SI{100}{MHz}$.

In conclusion, for high enough efficiency and phase control where the gate noise decrease persistently with increasing pump power instead of saturating, the gate noise in the cluster state computation scheme here can be brought down and eventually allow fault-tolerant computation, depending on the squeezing threshold set by the concatenated qubit error correction code. Besides this, the squeezing source bandwidth should be broad enough to cover the temporal modes in the frequency domain well. This latter condition is especially important to keep in mind when scaling up the number of encoded modes by reducing the short delay line, $\tau$, shortening the temporal modes in time domain and thereby broadening them in frequency domain.
